\documentclass[12pt,preprint]{aastex}
\usepackage[usenames]{color}
\newcommand{\eg}{{\it e.g.}\ }
\newcommand{\ie}{{\it i.e.}\ }
\newcommand{\etal}{{\it et al.}\ }
\newcommand{\myfontsize}{\fontsize{9}{11}\selectfont}
\def\gae{\lower 2pt \hbox{$\, \buildrel {\scriptstyle >}\over {\scriptstyle \sim}\,$}}
\def\lae{\lower 2pt \hbox{$\, \buildrel {\scriptstyle <}\over {\scriptstyle \sim}\,$}}

\begin{document}
\begin{center}
\vskip 8cm

{\bf Using the youngest asteroid clusters to constrain\\
the Space Weathering and Gardening rate on S-complex asteroids}

\vskip 1.4cm

{Mark Willman$^1$, Robert Jedicke$^1$, Nicholas Moskovitz$^2$, \\David
Nesvorn\'{y}$^3$, David Vokrouhlick\'{y}$^4$, Thais Moth\'{e}-Diniz}$^5$

\vskip 1.4cm

{$^1$Institute for Astronomy,  University of Hawai`i at Manoa\\
2680 Woodlawn Drive, Honolulu, HI{\ \ }96822}\\
{willman@ifa.hawaii.edu, 808-956-6989 tel, 808-956-9580 fax\\
jedicke@ifa.hawaii.edu, 808-956-9841 tel, 808-956-9580 fax\\

\vskip 0.5 cm

{$^2$Department of Terrestrial Magnetism\\
5241 Broad Branch Road, NW, Washington, DC 20008}\\
astromosk@gmail.com, 202-478-8862 tel}

\vskip 0.5 cm

{$^3$Department of Space Studies, Southwest Research Institute\\
1050 Walnut Street, Suite 400, Boulder, CO{\ \ }80302}\\
{davidn@boulder.swri.edu, 303-546-0023 tel, 303-546-9687 fax}

\vskip 0.5 cm

{$^4$Institute of Astronomy, Charles University\\
V Hole\v{s}ovi\v{c}k\'{a}ch 2, CZ-18000 Prague 8, Czech Republic}\\
{vokrouhl@cesnet.cz}
                                   
\vskip 0.5 cm

{$^5$UFRJ / Observat\'{o}rio do Valongo\\
Ladeira Pedro Ant\^{o}nio, 43\\
20080-090 Rio de Janeiro/RJ}\\ 
Brasil\\
{thais.mothe@astro.ufrj.br}\\

\end{center}

\vskip 0.0cm

\noindent 47 pages
\vskip 0.1cm

\noindent 7 figures

\vskip 0.1 cm

\noindent 4 tables

\clearpage

Running Head:  Space weathering and gardening rates on S-complex asteroids

\vskip 2.cm

Editorial correspondence to:

Mark Willman

Institute for Astronomy University of Hawaii'i at Manoa

2680 Woodlawn Drive, Honolulu, HI 96822

willman@ifa.hawaii.edu

808-956-6989 tel

808-956-9580 fax

\clearpage

\begin{center}
{\bf ABSTRACT}
\end{center}

We have extended our earlier work on space weathering of the youngest
S-complex asteroid families to include results from asteroid clusters
with ages $< 10^6$ years and to newly identified asteroid pairs with
ages $< 5 \times 10^5$ years.  We have identified three S-complex
asteroid clusters amongst the set of clusters with ages in the range
$10^{5-6}$ years --- (1270)~Datura, (21509)~Lucascavin and (16598)
1992 YC2.  The average color of the objects in these clusters agrees
with the color predicted by the space weathering model of
\citet{bib.wil08}.  SDSS 5-filter photometry of the members of the
very young asteroid pairs with ages $< 10^5$ years was used to
determine their taxonomic classification.  Their types are consistent
with the background population near each object.  The average color of
the S-complex pairs is $PC_1 = 0.49 \pm 0.03$, over 5$\sigma$ redder
than predicted by \citet{bib.wil08}.  This may indicate that the most
likely pair formation mechanism is a gentle separation due to YORP
spin-up leaving much of the aged and reddened surface undisturbed.  If
this is the case then our color measurement allows us to set an upper
limit of $\sim 64\%$ on the amount of surface disturbed in the
separation process.  Using pre-existing color data and our new results
for the youngest S-complex asteroid clusters we have extended our
space weather model to explicitly include the effects of regolith
gardening and fit separate weathering and gardening characteristic
timescales of $\tau_w = 960 \pm 160$ My and $\tau_g = 2000 \pm 290$~My
respectively.  The first principal component color for fresh S-complex
material is $PC_1 =0.37 \pm 0.01$ while the maximum amount of local
reddening is $\Delta PC_1 = 0.33 \pm 0.06$.  Our first-ever
determination of the gardening time is in stark contrast to our
calculated gardening time of $\tau_g \sim 270$~My based on main belt
impact rates and reasonable assumptions about crater and ejecta
blanket sizes.  A possible resolution for the discrepancy is through a
`honeycomb' mechanism in which the surface regolith structure absorbs
small impactors without producing significant ejecta.  This mechanism
could also account for the paucity of small craters on (433)~Eros.

\vskip 1cm



\section{Introduction}
\label{s.intro}

After three decades of development our understanding of the space
weathering phenomenon may be converging on its cause and effects.
Originally proposed \citep{bib.cha73} as a solution to the mismatch
between the color of the most common meteorites, ordinary chondrites,
and their likely source region, inner main belt S-complex asteroids,
the space weathering hypothesis is that the surface colors of
asteroids change with exposure to the space environment
\citep{bib.hap73,bib.hap68}.  The idea has held up to the test of time
and in recent years has led to measurements of the rate of color
change on S-complex asteroids yielding empirical models that predict
the rate of reddening of their surfaces \citep{bib.jed04,
bib.nes05,bib.wil08}.  This work extends our space weathering model to
the youngest dated asteroids in the main belt that are less than
$10^6$ years old and, for the first time, measures the surface
gardening rate on the asteroids due to impact-generated regolith
cycling.

\citet{bib.wil08}'s space weathering model applies exclusively to
S-complex asteroids.  Other types of asteroids may also undergo space
weathering and we expect that they would obey a model of similar
functional form with different parameters.  For instance, if C-complex
asteroids undergo space weathering they must have a different color
range because they occupy a different region of principal component
space and the weathering rate and even the sense of coloration may be
different \citep{bib.nes05}.  The gardening rate would probably also
be different due to different material density and strength.

In this work we study the space weathering of S-complex asteroids
without particular concern for the agent causing the weathering though
the likely culprit is energy deposition due to particle bombardment
\eg micrometeorites, solar protons and cosmic rays.  \citet{bib.mar06}
claim that Sun-related effects are dominant but they assumed that
micrometeorite impacts are independent of heliocentric distance.
Since neither their flux nor their velocity are distance independent
\citep{bib.cin92} it is unclear whether it is justified to downgrade
their contribution to the space weathering agent inventory.
Laboratory-based studies suggest that the asteroidal space weathering
mechanism, long confused with the lunar processes producing
agglutinates, involves the coating of near surface semi-transparent
nanophase grains by a vapor or sputter deposited film of metallic iron
\citep[\eg][]{bib.sas01,bib.pie00}.

While measuring the color of asteroid surfaces is relatively
straightforward and asteroid taxonomy based on colors has been a
mainstay of planetary science for decades the determination of
asteroid surface ages is a relatively recent innovation.  The surface
ages of asteroid family members may be determined from dynamical
simulations of the evolution of the family's orbit element
distribution\citep[\eg][]{bib.vok06a, bib.vok06b, bib.nes05,
bib.nes02, bib.mar95}.  The dynamical methods include family size
frequency distribution (SFD) modeling, global main belt SFD modeling,
modeling of family spreading via thermal forces and backward numerical
integration of orbits.  The combination of asteroid surface colors
with their ages allowed the first astronomical determination of the
space weathering rate \citep{bib.jed04,bib.par08}.

Space weathering rate measurements utilize the fact that asteroid
family members are fragments of a single collisionally disrupted
parent body that formed at the same time and assume that fresh
unweathered material from the collision debris cloud coats all members
of a family with a homogeneous regolith layer.  Therefore, all family
members should start out with similar color and weather in tandem.

We ignore grain size color effects for a number of reasons.  First, we
believe that unless grain size is correlated with age its effect will
average out over the members of a family.  Second, we will see below
that km scale asteroids, in contrast to moon sized bodies with highly
comminuted regoliths, are gravitationally sorted rubble
piles whose surfaces are dominated by boulders.  Fines are largely
absent, perhaps sequestered beneath the bouldered surface, so asteroid
regolith appears to vary with size.  This would diminish grain size
observational effects in the sub-milligee environment of small asteroids.
Finally, \citet{bib.nes05} and \citet{bib.jed04} did not find any
correlation between color and size, just color and age.

\citet{bib.jed04}'s space weathering rate measurements encapsulated
`the color' of an asteroid's surface as its first principal component
color
\begin{equation}
PC_1 = 0.396\;(u-g) + 0.553\;(g-r) + 0.567\;(g-i) + 0.465\;(g-z)
\label{eq.pc1}
\end{equation}
\noindent that correlates strongly with the average slope of the
spectrum.  The principal component color provides the linear
combination of filter magnitudes having the greatest variability over
the sample of asteroids.  In this case $PC_1$ is specific to the
asteroids in the Sloan Digital Sky Survey (SDSS) 2$^{nd}$ Moving
Object Catalog.  While most of the color variability in the SDSS
asteroid sample is captured in $PC_1$, the second principal component,
\begin{equation}
PC_2 = -0.819\;(u-g) + 0.017\;(g-r) + 0.090\;(g-i) + 0.567\;(g-z),
\label{eq.pc2}
\end{equation}
\citep{bib.nes05} corresponds to the asteroid spectrum's curvature and
is used in this analysis to identify asteroid taxonomy.

Motivated by the fact that uniform surface irradiation should produce
an exponentially decaying amount of unweathered surface
\citet{bib.jed04} showed that the color of S-complex asteroids as a
function of time $t$ can be expressed as
\begin{equation}
PC_1(t) = PC_1(0) + \Delta PC_1[1 - e^{-(t/\tau)^\alpha}]
\label{eq.jedpc1}
\end{equation}
\noindent where $PC_1(0)$ is the unweathered color of fresh surface
material, $\Delta PC_1$ is the magnitude of the weathering color
change after a long period of time, $\tau$ is the characteristic time
for space weathering and $\alpha$ is a generalizing exponent.
\citet{bib.wil08} improved upon the earlier work by refining the color
of the $\lae 5$~My old Iannini family, eliminating the Eos family that
is no longer considered to be in the S-complex, and carefully
refitting the data to the functional form given above to determine
$PC_1(0) = 0.31 \pm 0.04$, $\Delta PC_1 = 0.31 \pm 0.07$, $\tau = 570
\pm 220$ My and $\alpha = 0.53 \pm 0.19$.

The characteristic time scale of $570 \pm 220$~My for color change in
the main asteroid belt \citep{bib.wil08} is in agreement with pulsed
laser experiments \citep{bib.sas01} on silicate pellets intended to
simulate micrometeorite bombardment.  They suggested a characteristic
time for weathering of 100 My at $1$~AU, equivalent to about 700 My in
the main belt assuming that the Sun is the source of the weathering
agent and a $r^{-2}$ dependence on its strength.  However,
\citep{bib.loe09} argue that \citet{bib.sas01} overestimated the space
weathering characteristic timescale due to an incorrect flux
calculation leading.  Basically, the time scale extrapolation from lab
results is complicated because it is unclear how to evaluate the
contribution from several different factors.  For instance, in the
main belt micrometeorite impact speeds will be lower ($\sim 5$ km/s)
than at 1 AU ($\sim 20$ km/sec) and their impact energy will be
$\sim$16x weaker.

\citet{bib.pie00} estimated characteristic aging times of $100-800$ My
for lunar surfaces by comparing ages from craters dated
radiometrically or by cosmic ray exposure ages to spectral
differences.  Correcting the rate to the center of the main belt
suggests weathering times on the order of $\sim$600-4800~My.  Even
though the lunar surface is not identical to S-complex asteroid
surfaces the time scales are similar.

Finally, we note that craters on (243)~Ida are bluer than their
surrounding background terrain \citep{bib.vev96}.  The craters
correspond to freshly exposed and unweathered regolith
\citep{bib.vev96} while other parts of the asteroid's surface
indicates an age of about 1 Gy \citep{bib.gre96}.  The wide range in
diameters (a proxy for crater age) of blueish crater suggests that the
space weathering time must be long.

On the other hand there are claims of faster surface weathering time
scales such as \cite{bib.tak08}'s upper limit of 450 kyr based on a
shallow 1 $\mu$m absorption band observed on (1270)~Datura.  However,
we expect that the space weathering phenomenon is a relatively subtle
effect easily masked by stochastic variations between asteroids due to
\ie mineralogical and morphological differences and/or collisional and
cratering events.  The space weathering effect can only be identified
in specific regions on an asteroid as observed with {\it in situ}
spacecraft measurements on (951)~Gaspra \citep{bib.hel94}, (243)~Ida
\citep{bib.vev96}, and (25143)~Itokawa \citep{bib.ish07}] or as an
  ensemble effect on a statistically large sample of asteroids.  It is
  therefore difficult to make a general conclusion based on a single
  asteroid such as (1270) Datura.  Indeed, our measurement of that
  asteroid's $PC_1 = 0.41 \pm 0.02$ is redder by $2.2 \sigma$ than its
  predicted color from \citet{bib.wil08}.  Takato's measurement of a
  shallow 1 $\mu$m absorption band which would tend to redden the
  overall spectrum is therefore generally consistent with our
  measurement.  We expect that the `redness' of individual small
  rubble pile asteroids in a sample is affected more by random surface
  variations than larger regolith-rich asteroids with finer surface
  materials.  In this case (1270) Datura is redder than the Datura
  cluster average so its particular value can be misleading.  The four
  (1270) Datura members in this work have mean $PC_1 = 0.305$ with a
  RMS of $0.278$.  The large standard deviation is not due to
  measurement error --- it is a result of intrinsic color differences
  between the members of the (1270) Datura family and is typical of
  other families.  Therefore we would discount the resulting 450 kyr
  weathering time.

Another short time scale measurement was reported from recent lab
experiments on olivine powder by \citet{bib.loe09} (and references
therein) simulating solar wind effects at 1 AU by bombardment with 4
keV protons.  They conclude that spectral reddening caused by the
solar wind should saturate in $\sim$ 5 kyr.

\citep{bib.ver09} combined spectral data from a sample of four members
of young clusters with archival meteorite spectra and found that space
weathering is substantially complete in $<$ 1 My but then continues at
a slower pace up to several Gy.  They attribute the first stage to the
solar wind (\citet{bib.str05} lab simulation) and the second to
micrometeorite bombardment.  This scenario depends critically on the
starting color that they derive from meteorite data.

It is not clear how to reconcile the information from these various
lab results that differ in time scale by several orders of magnitude
with our observations.  

\citet{bib.wil08}'s weathering model was derived from families that
were several~My (\eg Karin, Iannini) to several Gy (\eg Eunomia,
Maria) old.  The family age estimates have typical errors of $\sim
40$\% resulting from fundamental limitations in the dating techniques
\citep{bib.nes07}.

Although there were nine known S-complex families with ages from tens
of Mys to a few Gys at that time there were only two younger than 10
My.  Refining the space weathering rate at even younger ages requires
a large sample of young asteroid families, which are typically
produced by the catastrophic disruption of small asteroids, have only
a small number of detectable members, and are therefore difficult to
identify.  But in recent years the number of catalogued asteroid
orbits has reached hundreds of thousands and includes many asteroids
smaller than 1~km.  This large sample of asteroids allows the
identification of rare small clusters\footnote{As pairs of asteroids
  are (tautologically) composed of two asteroids, the established
  families have dozens to thousands of members, and known `clusters'
  have between three and seven members we define a cluster as a small
  family with three to about ten asteroids.  Pairs have distinct
  formation mechanisms from families and clusters.  Families form from
  larger parent bodies than clusters and are therefore older on
  average.  The youngest known family, Iannini, is $\sim$3 My old
  while the clusters are less than 1 My old.} originating from
collisions less than 1 My ago.  \citet{bib.nes06a,bib.nes06b} found
four such clusters with a total of 16 members.  The clusters are named
after their largest known members: (1270)~Datura,
(14627)~Emilkowalski, (21509)~Lucascavin and (16598) 1992 YC2.

The progression to ever smaller clusters reached its logical limit
with \citet{bib.vok08}'s discovery of 60 pairs of asteroids in
extremely similar orbits --- much more alike than would be expected
based on the density of proper elements for other asteroids with
similar orbit elements.  The pairs are thought to have formed less
than 500 kyr ago based on the dynamical evolution time of the pair
member's orbits.  Their list was updated and restricted to only 36
pairs \citep{bib.pra09b}.  Pairs belonging to known young families
such as (1270) Datura or (832) Karin were excluded because of the
possibility that the catastrophic impact and subsequent family
formation process may create paired asteroids.  Such cases were
eliminated to focus on pairs that were created in isolation and
presumably by the same method.  The formation method of pairs may
involve a critical distinction from that of clusters or families.
Whereas the latter form in catastrophic collisions, pairs have
alternative possible formation methods.  Some of these methods may be
gentle processes not involving resetting of the surface as discussed
in \S\ref{ss.comparemodel}.

Shortly after the publication of \citet{bib.wil08} we obtained
observations of some members of the sub-My old families and found that
they were significantly redder than predicted.  Did this imply a
problem with our space weathering model or is the space weathering
phenomenon more complicated than suggested by the simple model of
eq.~\ref{eq.jedpc1}?

We were also concerned with the functional form of the space
weathering model of \citet{bib.wil08} \ie eq. \ref{eq.jedpc1}.  If
space weathering is an isolated process the exponent $\alpha$ should
be unity but \citet{bib.jed04} fit $\alpha = 0.53 \pm 0.19$.  Their
argument was that $\alpha$ is a generalizing factor that accounts for
both space weathering and regolith gardening which acts to counteract
the surface aging by slowing turning over the asteroid's surface.

To investigate these questions we collected color and spectral data
from several sources for members of the young families and pairs.  We
obtained spectra of members of the sub-My asteroid clusters
\citep{bib.nes06a,bib.nes06b} and were provided spectra for some of
the objects observed by \citet{bib.mot08}.  We also located archived
photometry for 19 pair members in the Sloan Digital Sky Survey Data
Release 7 Moving Object Catalog 4 (SDSS DR7 MOC4) \citep{bib.par08}.
We then developed a new asteroid surface color-age model that
explicitly separates the effects of weathering and gardening and
eliminates the need for the unexplained generalizing factor
($\alpha$).  The new model allowed us to measure the characteristic
timescales for both weathering and gardening on the S-complex
asteroids.  Finally, we independently calculated the gardening
timescale on main belt asteroids from their size distribution, impact
rates and estimates of crater and ejecta blanket size.

\section{Space weathering versus regolith gardening effects}
\label{s.weathergarden}

The space weathering model of \citet{bib.jed04} summarized in
eq. \ref{eq.jedpc1} uses a characteristic time, $\tau$, along with a
generalizing and unphysical exponent, $\alpha$, to capture the time
dependence of the gradual reddening of S-complex asteroid surfaces.
The exponent was introduced because it was understood that there are
more effects in play on an asteroid's surface than a single weathering
component.  \eg multiple sources of space weathering with different
time scales such as solar protons and ultraviolet radiation,
micrometeorite bombardment and cosmic rays, in addition to the effect
of regolith gardening which will counteract the space weathering.
`Gardening' is in part due to meteoroids that regularly strike the
asteroids's surface and lift fresh sub-surface material to the
regolith's top layer.  Gardening may also take place through seismic
shaking with subsequent regolith distribution or it may result from
larger asteroid strikes which spread ejecta blankets beyond a crater.
We do not distinguish between possible causes but lump them all under
the banner of 'gardening' in the same way that all possible causes of
color change in minerals are included in 'weathering'.  Here we extend
the space weathering model of \citet{bib.jed04} to explicitly include
both weathering and gardening.

Consider the relationship of unweathered surface, $U$, and weathered
surface, $W$ --- by construction $U+W=1$.  Space weathering causes $U$
to be replaced by $W$ at the rate
\begin{equation}
\frac{dU^{-}}{dt} = -\frac{U(t)}{\tau_w}
\label{eq.unweathered_surface}
\end{equation}
\noindent where $\tau_w$ is the characteristic space weathering time.
Regolith gardening causes $U$ to increase at the same rate that $W$
decreases,
\begin{equation}
\frac{dU^{+}}{dt} \equiv - \frac{dW}{dt} = \frac{W(t)}{\tau_g}
\end{equation}
\noindent where $\tau_g$ is the characteristic gardening time.  

The total rate of change of unweathered surface is then
\begin{equation}
\frac{dU}{dt} = \frac{dU^{+}}{dt} + \frac{dU^{-}}{dt} = \frac{1}{\tau_g} - \left(\frac{1}{\tau_g} + \frac{1}{\tau_w}\right)U.
\label{eq.unweathered}
\end{equation}
\noindent Since $U(0) = 1$
\begin{equation}
U(\tau_g,\tau_w,t) = \frac{e^{-\left(\frac{1}{\tau_g} + \frac{1}{\tau_w}\right)t} + \frac{\tau_w}{\tau_g}} {1 + \frac{\tau_w}{\tau_g}}
\label{eq.unweatheredsoln}
\end{equation}
\noindent which has the desired properties that $U(t=0)=1$ and as $t
\rightarrow \infty$, $U \rightarrow (1+\frac{\tau_g}{\tau_w})^{-1}$
--- the amount of unweathered asteroid surface after a long period of
time is related to the relative rates of space weathering and regolith
gardening.

Replacing the unweathered surface term in eq.~\ref{eq.jedpc1} with our
new generalized unweathered surface term in
eq.~\ref{eq.unweatheredsoln} yields
\begin{equation}
PC_1(t) = PC_1(0) + \Delta PC_1[1 - U(\tau_w,\tau_g,t)]
\label{eq.garden}
\end{equation}
\noindent which includes the effects of both weathering and gardening
and will allow the two characteristic times to be separately
determined by fitting to the asteroid families' color and age data.

Previous attempts to explore the relationship between weathered and
unweathered surface have foundered on the fundamental equations.  For
instance, \citet{bib.gau74} overlooked the restorative effects of
regolith gardening and assumed\footnote{\citet{bib.gau74} eq. 1 is
analogous to our eq. 4.} that the fraction of surface that is
undisturbed is simply a decaying exponential in time.

We return to the model parameters that best fit the color-age data in
\S\ref{ss.comparemodel}.  That gardening time determined from
age-color data is compared to the resurfacing time determined from
impacts in \S \ref{s.gdntime}.  In the next section we calculate the
resurfacing rate using the independent method of impactors.

\section{Calculating the resurfacing rate from impacts}
\label{s.gardenimpact}

The asteroid resurfacing rate due to impacts, $\dot{S}$, is the
product of the frequency of impacts, $\nu(D)$, and the ejecta covered
surface area, $A_e(D)$ per impact, where both terms are a function of
impactor diameter $D$.  Initially we place no limit on diameter which
could extend from micrometeoroids up to parent body shattering size.
The constraint on size will enter the discussion below in the lower
limit of the impact integral.  We realize that the asteroid's surface
may also be indirectly affected by impact-induced seismic shaking
\citep{bib.ric05} which would increase the resurfacing rate.  Seismic
shaking could even happen without impact.  \citet{bib.bin10} find that
near Earth objects (NEOs) with likely close Earth encounters within
the past 500 kyr have bluer surfaces than their counterparts that
avoid such encounters suggesting that seismic shaking induced in this
manner is effective at resurfacing.  However, the mechanism of
resurfacing during close encounters with massive planets is not
important for MB asteroids.

On the other hand, since asteroids in the 1---10 km diameter range
have escape velocities of $\sim$1---10 m/s it is possible that a
sizable fraction of ejecta on those asteroids is lost to space which
would have the effect of decreasing the resurfacing rate.

The impact frequency is the product of the differential size frequency
distribution, $N(D)$, the impact cross section,
\begin{equation}
c(D_T,D) = \biggl( \frac{D_T+D}{2} \biggr)^2
\label{eq.crosssection}
\end{equation}
\noindent and the intrinsic collision probability, $P_i = 2.86 \times
10^{-18}$~km$^{-2}$~yr$^{-1}$ \citep{bib.bot94} --- the main belt wide
average collision probability that a single member of the impacting
population will hit a unit area of the target body per unit time.
$D_T$ is the target's diameter and the usual factor of $\pi$ in
eq. \ref{eq.crosssection} is implicitly included in $P_i$.  The cross
section is determined by the limiting distance between centers at
which the edge of the impacting asteroid just glances the edge of the
target asteroid: $\frac{D_T + D}{2}$.

Combining the above and summing over all impactor sizes up to the
maximum diameter ($D_{max}$) that will not collisionally disrupt the
target asteroid yields the surface gardening rate on the target
asteroid:
\begin{equation}
 \frac{dS(D_T)}{dt} = P_i \; \int_{\sim 0}^{D_{max}} \; N(D)
 \; c(D_T,D) \; A_e(D) \; dD \; .
\label{eq.resurfacerate}
\end{equation}
The resurfacing rate per unit target area is then $\dot{S}(D_T) / \pi
D_T^2$.  The lower limit on the integral is $\sim 0$ because small
dust-size particles get blown away by solar radiation pressure and
`dust' up to about 1~cm diameter spirals into the Sun due to the
Poynting-Robertson effect \citep{bib.der02}.  The size-frequency
distribution used here \citep{bib.bot05b} is only specified down to
$\sim 1$~m diameter but we will show later that the size range from
1~cm to 1~m does not affect our conclusions.

Since gardening of already gardened surface has no effect we are only
interested in the fraction of gardening that takes place on weathered
surface.  Using the notation of \S\ref{s.weathergarden} the rate
of gardening on weathered surface is:
\begin{equation}
\frac{dU^{+}}{dt} = W \; \frac{\dot{S}(D_T)}{\pi D_T^2}
\label{eq.fracgdn}
\end{equation}
leading to an exponentially decreasing amount of weathered surface
with characteristic time
\begin{equation}
\tau_g = \frac{\pi \; D_T^2}{\dot{S}(D_T)}
\label{eq.gdntime}
\end{equation}

Before combining equations \ref{eq.crosssection} $-$ \ref{eq.gdntime}
into final form there are several additional factors to consider that
are covered in the following sub-sections: \S\ref{ss.largestImpactor}
the diameter of the largest impactor, $D_{max}$, that will not shatter
the target asteroid, \S\ref{ss.smallestImpactor} the diameter of the
smallest impactor, $D_{min}$, that will produce a significant ejecta
blanket, \S\ref{ss.craterRadius} the diameter, $D_c$, of the crater
produced by an impactor of diameter, $D$, and \S\ref{ss.ejectaRadius}
the diameter, $D_e$, of the ejecta blanket.  We finally calculate the
regolith gardening rate in \S\ref{ss.resurfacingTime}.

\subsection{Largest non-shattering impactor, $D_{max}$}
\label{ss.largestImpactor}

The size of an impactor that will shatter a target is determined by
the shattering specific impact energy, $Q_S^*(D_T)$.  This is
different from, and smaller than, the catastrophic disruption specific
impact energy, $Q_D^*(D_T)$, that applies to shattering the target and
dispersing the fragments with high enough energy that reassembly into
another rubble pile is impossible.  Since we are concerned with
regolith gardening we require that some portion of the asteroid's
surface remain intact and use $Q_S^*(D_T)$ to determine the diameter
of the largest non-shattering impactor.  Higher energy could, at a
minimum, reset the entire asteroid surface to one with no weathering
history or fission the target body.  In either case the new surface
would start with no history and would be unidentifiable with the
previous surface.

Smaller, typically monolithic objects are relatively resistant to
catastrophic disruption and, counterintuitively, large gravitational
aggregates (rubble piles) have high collision strength because of
their ability to absorb energy through non-elastic compression.  While
there are a wide variety of both theoretical and empirical
$Q_S^*(D_T)$ functions the specific energy is typically at a minimum
for objects with diameters in the range $0.1-10$~km
\citep[\eg][]{bib.hol02}.

It can be shown that the ratio of impactor to target diameters
required to shatter small asteroids is
\begin{equation}
\frac{D_{max}}{D_T} = \biggl( \frac{2Q_S^*}{v^2}\frac{\rho_T}{\rho_I} \biggr)^{1/3}
\label{eq.Dmax}
\end{equation}
\noindent where $v$ is the collision speed, $\rho$ represents density
 and the sub-scripts $I$ and $T$ represent the impactor's and target's
 quantities respectively.  The impactor and target densities are
 assumed to be equal as the most likely scenario in the inner main
 belt where our sample is located involves collisions between two
 S-complex asteroids of any size.

Using \citet{bib.hol02}'s $Q_S^*(D_T)$ and
 $v=5$~km/s typical of main belt asteroid collision speeds we find
 that $D_{max}$:$D_T \sim $1:40 for asteroids of a few kms diameter
 like those in our data.

\subsection{Smallest ejecta blanket-creating impactor, $D_{min}$}
\label{ss.smallestImpactor}

We will show below that the calculated gardening rate is strongly
dependent on the size of the smallest impactor capable of creating an
ejecta blanket or, more generally, affecting an area substantially
larger than its own cross section.  Basically, because there are a
large number of small impactors their contribution to the gardening
rate could be substantial if not limited in same way by the target
asteroid's structure or composition.  For instance, if an asteroid's
surface is mostly covered by meter-scale boulders it might require
$\gg$-meter scale impactors to generate a significant gardening
signature.

The microgravity environment on small asteroids ($\la$ 1 km diameter)
displays surprising phenomena as illustrated by images of
(25143)~Itokawa \citep{bib.miy07} obtained by the Hayabusa mission
\citep{bib.fuj06}.  The asteroid's surface is imbricated in many
areas --- cobble-sized stones coat the surface in a fairly regular
pattern and are oriented in a common direction in the manner of
bricks.  Fine materials are largely absent from the surface except
where they are collected in low gravity potential ponds.  The fines
may have sifted downwards through the larger rocks \citep{bib.miy07}
in addition to being preferentially dispersed into space during
impacts \citep{bib.cha78}.  This rubble pile of gravel appears to have
been size sorted into an inside out hierarchy; the largest rocks being
at the surface and fines hidden inside.

We think that an imbricated surface may act akin to medieval chainmail
effectively warding off blows and protecting the surface from impact
cratering.  Evidence supporting this mechanism may be found in
\citet{bib.miy07} Figures 2 e,f where light colored ejecta from the
adjacent Komaba crater appears to coat one side of scattered bricks
but there is no continuous fine coating of the surface beyond the
crater rim.  This suggests that the coating was not broadcast over the
landscape as a powder but rather that these rocks were thrown from the
crater to their current position carrying the coating with them.  This
would be the result if the coated rocks had originally overlain the
position of an impact which caused them to be ejected with a dust
coating on their undersides.  The key point is that the dust itself
was inhibited from being broadcast to resurface a wide area.  We
therefore postulate a `chainmail' mechanism that results in resistance
to penetration and ejecta creation by impactors smaller than some
multiple of the brick size.  The mechanism was termed `armoring' by
\citet{bib.cha02} but we believe `chainmail' is a slightly more
descriptive term for the proposed surface quality.  Armor brings to
mind rigid continuous sheets rather than the flexible linked elements
in chainmail.  The `chainmail' mechanism may have a significant
influence on the effect of regolith gardening on small asteroids since
small meteoroid impacts will be relatively inefficient at gardening
compared to the ejecta blankets produced by larger impactors.

With this in mind we allow that impactors create craters according to
accepted crater scaling laws for $D>D_{min}$ \citep{bib.mel89}.  For
smaller impactors, those in a size range corresponding roughly to the
size of the imbricated surface regolith, we will set the diameter of
the affected area to $D$.  But what is the appropriate value for
$D_{min}$?

The only two asteroids that have been imaged at a surface resolution
better than $\sim$50 m are (433) Eros and (25143) Itokawa.  Both show
imbricated surfaces with only small areas covered by fines.  Figure 4b
of \citet{bib.miy07} provides the cumulative size frequency
distribution of gravel of diameter $d$ on Itokawa with $N \propto
d^{-2.8}$ but there is a bump near $d$ = 2 m which is confirmed by
visual inspection of the boulder strewn areas and a broad hump near
$d$ = 0.3 m.  The smooth areas dominated by finer material are the
rare exceptions on (25143) Itokawa.  For (433)~Eros, Figure 3 from
\citet{bib.cha05} shows a surface that is 'bumpy' near the resolution
limit of 1 m which we think implies that the surface elements are
$\sim$1 m in size.

With an ignorant assumption that the impactor needs to be $\ga
10\times$ more massive than the surface elements to be large enough to
apply the crater scaling laws it would imply that $D_{min} \sim 2$~m.
For the purpose of this work we scale that figure up by another order
of magnitude and take $D_{min} = 4$~m as our nominal case.  We will
show in the discussion that using smaller $D_{min}$ only serves to
increase the discrepancy between our measured and calculated gardening
rates so that the choice of the nominal value is unimportant.

\subsection{Crater diameter, $D_c$}
\label{ss.craterRadius}

While impact craters on Earth typically have diameters $D_c \sim 10
\times D$ \citep{bib.dep01} this relationship should probably not be
be used for the low gravity high-porosity cratering impacts taking
place between asteroids in the main belt.  In this sub-section we show
that we can not use standard crater scaling laws because they appear
to over-estimate the $D_c/D$ ratio on asteroids and we will argue that
it makes more sense to use a $D_c/D$ ratio that is only weakly
$D$-dependent.

The general empirical scaling law\footnote{We repeated the calculation
  with a similar equation from \citet{bib.sch87} having slightly
  different exponents which apply to porous media.  The result did not
  alter our conclusions.} eq. 5.26b of \citet{bib.dep01} for crater
  diameter is (in mks units)
\begin{equation}
D_c = 1.8 \; 
      \rho_I^{0.11} \; 
      \rho_T^{-1/3} \; 
      g_T^{-0.22} \; 
      E_k^{0.22} \; 
      \sin^{1/3} \theta \;
      D^{0.13}
\label{eq.craterradius}
\end{equation}
\noindent where $g_T$ is the gravitational acceleration on the
target's surface, $E_k$ is the kinectic impact energy, and $\theta$ is
the angle of impact from the local horizontal.  Since the impact
energy, $E_k$, is expressed in terms of $D$, $\rho_I$ and the impact
speed $v$ then
\begin{equation}
D_c = 1.34 \; 
      \biggl( \frac{\rho_I}{\rho_T} \biggr)^{1/3} \; 
      g_T^{-0.22} \; 
      v^{0.44} \; 
      \sin^{1/3} \theta \;
      D^{0.79}.
\label{eq.craterradiusvel}
\end{equation}
Ignoring the gravitational focussing on the Earth\footnote{We will
  show in the discussion that the systematic errors are larger than
  the amplitude of this effect.}, the ratio of crater diameters on an
  asteroid and the Earth for the same size impactor is
\begin{equation}
\frac{D_c(\rm{asteroid})}{D_c(\rm{Earth})} 
= 
\biggl( \frac{\rho_E}{\rho_{A}} \biggr)^{1/3}
\biggl(\frac{g_E}{g_{A}} \biggr)^{0.22}
\biggl (\frac{v_A}{v_E} \biggr)^{0.44}
\label{eq.asteroidEarthDcRatio}
\end{equation}
But can this result be used to determine the ratio over the three
orders of magnitude difference between the size of the Earth and the
asteroids in our sample?

Using $\rho = 2.7$ g/cm$^3$ with $g = 0.46$ cm/s$^2$ for (433)~Eros
\citep{bib.kor04} and $\rho = 2.6 \pm 0.5 $ g/cm$^3$ \citep{bib.bel95}
with $g = 1 $ cm/s$^2$ \citep{bib.kor04} for (243)~Ida yields a simple
arithmetic mean target density for the two asteroids of 2.65 g/cm$^3$
and gravity of 0.73 cm/s$^2$.  Assuming an impactor density equal to
the average S-complex asteroid of 2.63 g cm$^{-3}$
\citep{bib.hil02,bib.fuj06}, a density of 2.7 g cm$^{-3}$
\citep{bib.col05} for the upper layer of the Earth's continental
crust, typical impact speeds on the Earth and asteroid of $v\sim$~20
km/s \citep{bib.col05} and $v =$~5~km/s \citep{bib.bot94}
respectively, we find that $D_c(\mathrm{asteroid}) /
D_c(\mathrm{Earth}) \sim 2.7$.  Then, since $D_c \sim 10 \times D$ on
the Earth \citep{bib.dep01} the scaling law leads to an analogous
ratio for two S-complex asteroids of $D_c \sim 27 \times D$.

However, \citet{bib.bot06d} have used crater counting methods and an
assumed impactor size distribution to estimate that $D_c \sim 12.5 \;
\times \; D$ on (433)~Eros and (243)~Ida.  So a decrease in
scale from the Earth's diameter to the 10s of km size of (243)~Ida
and (433)~Eros only increases the $D_c$:$D$ ratio from 10 to 12.5
in contrast to the scaling calculation in the last paragraph that
suggested a ratio of $\sim 27\times$.  Thus, it appears that the
scaling relation of eq. \ref{eq.craterradiusvel} fails to capture
essential physics involved in crater formation on small asteroids
although it is not surprising that an extrapolation from the Earth's
size down to kilometer scale fails.  We cannot use the
\citet{bib.dep01} equation or even the \citet{bib.sch87} equation
specific to porous media to calculate crater size on our few km scale
asteroids.

We empirically selected a power law to fit the two data points, Earth
with mean radius 6371 km and $D_c / D = 10$ and Eros \& Ida with mean
radius 16.3 km and $D_c / D = $ 12.5 yielding
\begin{equation}
\frac{D_c}{D} = 13.73 \; \biggl( \frac{D_T}{\mathrm{km}}\biggr)^{-0.03350}.
\label{eq.craterdiamsimple}
\end{equation}
\noindent The near zero exponent indicates the weak dependence of the
crater diameter ratio on target size and allows us to justify
extending the relationship to the few km diameter range of the
asteroids typical in our sample.


\subsection{Ejecta blanket diameter, $D_e$}
\label{ss.ejectaRadius}

The final key to calculating the asteroid regolith gardening rate from
impact rates and cratering effects is the area affected by the crater
ejecta.  Like all the other terms in the calculations, this effect is
difficult to characterize because cratering on small asteroids is not
as well studied as on planets and their satellites.  Due to their weak
gravity a portion of ejecta from typical impacts, particularly the
fine material \citep{bib.nak94}, may be thrown entirely clear of the
target resulting in reduced ejecta coverage \citep{bib.cha02}.

The fraction of material ejected from large craters is lower for
higher porosity objects \citep{bib.hou03}.  \citet{bib.bri06} estimate
macroporosities for coherent objects such as the moon, S-class
asteroids, and C-class asteroids to be $\lesssim 2-3 \%$, $15-20 \%$,
and $25-50 \%$ respectively.  Hence we would expect porosities of the
asteroids composing our sample of S-complex families to be some
$7\times$ higher than that of the Moon leading to smaller ejecta
blankets on asteroids compared to the Moon.

\citet{bib.mel89} describes continuous ejecta blankets as typically
coating the lunar surface out to over two crater rim radii from the
center.  His eq. 6.3.1 gives the radius of continuous ejecta, $R_e$,
as a nearly linear function of the crater rim radius, $R_c$, with $R_e
= (2.3 \pm 0.5) \, R_c^{1.006}$ for 1.3~km$<R_c<$436~km.  This
function roughly applies to Mercury, Mars, the Jovian satellites
Ganymede and Callisto, and the Saturnian satellites Dione and Rhea.

Bearing in mind the uncertain porosity and gravity effects we use $R_e
= (2.3 \pm 0.5) \, R_c^{1.006}$ as our baseline estimate and now
attempt to account for the distal rays. We imagine a torus immediately
outside the continuous ejecta disc with an area equivalent to
integrating the patchy distal ray coverage.  Using the lunar crater
Timocharis (see Figure 6.2 in \citet{bib.mel89}) as a prototypical
case we estimate that replacing the rays with such a torus increases
the continuous ejecta disc of $\sim 2.3 \, R_c$ to about $2.7 \, R_c$
yielding an area $\sim 38$\% larger than the continuous ejecta blanket
alone.  Converting radius to diameter and combining $D_e = (2.7 \pm
0.5) \, D_c^{1.006}$ and eq. \ref{eq.craterdiamsimple} yields $D_e =
(37.7 \pm 7.0) \; D^{1.006} \; D_T^{-0.0337} $ or $D_e \sim 38 \; D$.

Thus, the area covered by ejecta for $D > D_{min}$ is
\begin{equation}
A_e(D_T, D) = \frac{\pi}{4} \; D_e^2 = 1100 \; D_T^{-0.067} \; D^{2.01}.
\label{eq.blanketarea}
\end{equation}
For $D \le D_{min}$ we simply assume that the affected area is equal
to the impactor's cross section: $A_e(D_T, D) = \frac{\pi}{4} \; D^2$.

\subsection{Resurfacing time distribution}
\label{ss.resurfacingTime}

\begin{figure}[ht!]\small
\centerline{\includegraphics[width=5.0in,angle=90]{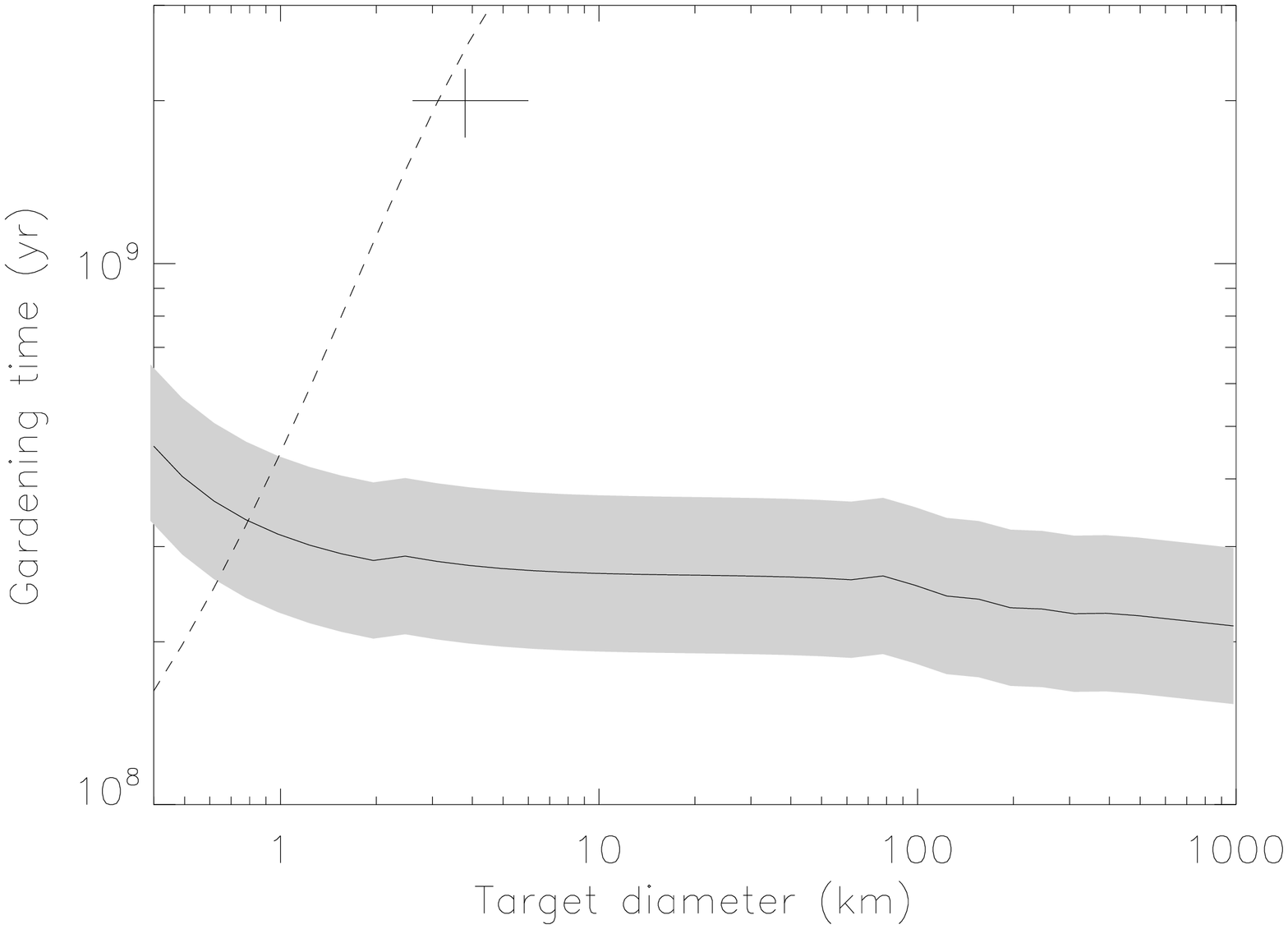}}
\caption\small{Calculated gardening time (solid curve) from asteroid
  impact and cratering rates as a function of target diameter.  The
  shaded region is our formal estimate of the statistical error on the
  gardening given the uncertainty on all the input parameters.  The
  actual systematic error is much larger as discussed in the text.
  Our measured color-derived gardening time is the single data point
  in the upper left.  The asteroid disruption lifetime is given by the
  dashed curve \citep{bib.bot05b}.  The histogram (right ordinate)
  shows the size distribution of all the asteroids used in this study
  assuming an albedo for the S-complex asteroids of 0.20.}
\label{f.gardentaulog}
\end{figure}

Combining all the discussion in this section above we find that the
characteristic gardening time from eq. \ref{eq.gdntime} is given by
\begin{equation}
\tau_g(N(D),D_T,Q_S^*,v,\rho_T,\rho_I) \; \gae \; \frac{9.8 \times
10^{14} \; D_T^{2.067}} {\int_{\sim
0}^{D_{max}(Q_S^*,v,\rho_T,\rho_I)} \; N(D) \; \biggl( \frac{D_T+D}{2}
\biggr)^2 \; D^{2.01} \; dD}.
\label{eq.tauofR}
\end{equation}

Figure \ref{f.gardentaulog} provides our nominal estimate of the
characteristic gardening time using $Q_S^*(D_T)$ from
\citet{bib.hol02}, $v = 5$ km/s \citep{bib.bot94}, $\rho_I = \rho_T =
$2.63 g cm$^{-3}$ \citep{bib.hil02,bib.fuj06} and $N(D)$ from
\citet{bib.bot05b}.  We find that the gardening time frame is a mild
function of diameter, only 23\% slower on small (1 km) asteroids than
on large (100 km).  The calculated resurfacing time for $D_T \sim
4$~km, the typical diameter of asteroids in our study, is $\sim 270$
My.  Note that asteroids with diameters $\lae$700~m have disruption
lifetimes shorter than either the calculated gardening time or the
weathering time implying that few of them would be fully weathered or
gardened.  Since (25143) Itokawa is $\sim$535~m long it is unlikely to
be completely reddened and \citet{bib.ish07} provide evidence of some
space weathering.  Unfortunately, the three bvw filter bands used
\emph{in situ} on (25143) Itokawa are not comparable to the five SDSS
ugriz filters and we cannot calculate $PC_1$ to determine where
(25143) Itokawa's color falls on our scale.

As new surveys detect asteroids of smaller diameters we may be able to
discover this transition size, further constraining the
weathering/gardening model.  However, the asteroids in our sample have
lifetimes $\gae 2$~Gy so their record of space weathering and
gardening should not be distorted by an early demise.

The above result for the resurfacing time determined by impacts is
compared to the gardening time determined via age-color data in
\S\ref{s.gdntime}.

\section{Data acquisition and reduction}\label{s.datareduct}

We observed or obtained spectra or photometry for ten asteroids within
the four known sub-My clusters as shown in Table~\ref{t.Sub-My Asteroid
Clusters}.  The data came from three sources: 1) new spectra and
photometry were acquired by our team using KeckII/ESI
\citep{bib.she02}, UH2.2~m/SNIFS \citep{bib.lan04} and IRTF/SpeX
\citep{bib.ray03}, 2) spectra were also provided by \citet{bib.mot08},
hereafter referred to as MDN, and 3) we identified photometry for some
sources from SDSS DR7 MOC4 \citep{bib.par08}.

\begin{table}[!ht]\small
\begin{center}
\title{Table~\ref{t.Sub-My Asteroid Clusters}. Sub-My Asteroid Clusters}
\begin{tabular}{ccccccc}
& & \\
\tableline\tableline
Asteroid             &Observation (HST)& Source  & Mag (H/V) & Type    & Axis  \\
\tableline
(1270) Datura        & 2007 Oct 28 & UH/SNIFS    & 12.5/17.1 & Sl      & 2.234 \\
(1270) Datura        & 2008 Mar 2  & KeckII/ESI  & 12.5/16.5 & Sl      & 2.234 \\
(1270) Datura        & 2007 Nov 16 & MDN         & 12.5/16.7 & Sk      & 2.234 \\
(90265) 2003 CL$_{5}$& 2007 Mar 15 & MDN         & 15.4/19.3 & Sq      & 2.235 \\
(60151) 1999 UZ$_{6}$& 2007 Mar 15 & MDN         & 16.1/20.5 & Sk      & 2.235 \\
(60151) 1999 UZ$_{6}$& 2001 Mar 18 & SDSS        & 16.1/20.1 & Sq/Sk   & 2.235 \\
(203370) 2001 WY$_{35}$& 2007 Sep 2 & MDN        & 17.6/21.5 & O/Q     & 2.235 \\
2003 UD$_{112}$      & 2003 Sep 25 & SDSS        & 18.4/20.0 & Sq      & 2.233 \\
\tableline
(14627) Emilkowalski & 2006 Oct 1  & UH/SNIFS    & 13.1/16.6 & T       & 2.598 \\
(14627) Emilkowalski & 2008 Apr 13 & UH/SNIFS    & 13.1/17.3 & T       & 2.598 \\
(14627) Emilkowalski & 2008 Apr 18 & IRTF/SpeX   & 13.1/17.4 & -       & 2.598 \\
(14627) Emilkowalski & 2004 Jan 27 & SDSS        & 13.1/18.3 & D       & 2.598 \\
\tableline
(16598) 1992 YC2     & 2007 Aug 17 & MDN         & 14.7/20.3 & Sq      & 2.621 \\
(16598) 1992 YC2     & 2000 Jan 1  & SDSS        & 14.7/17.0 & S       & 2.621 \\
\tableline
(21509) Lucascavin   & 2006 Aug 1  & UH/SNIFS    & 15.0/19.2 & Sk      & 2.281 \\
(21509) Lucascavin   & 2008 Apr 18 & IRTF/SpeX   & 15.0/18.1 & -       & 2.281 \\
(180255) 2003 VM$_{9}$& 2008 Mar 2 & KeckII/ESI  & 17.0/19.7 & Sk      & 2.280 \\
(209570) 2004 XL$_{40}$& 2007 Aug 20& MDN        & 17.1/20.4 & Sq      & 2.281 \\
\tableline\tableline
\end{tabular}
\end{center}
\caption\small{Observations and basic properties of members of four
  sub-My old asteroid clusters adapted from \citet{bib.nes06a} and
  \citet{bib.nes06b}.  The four clusters are separated by table lines
  and named after the first object listed in each cluster.  Dates of
  observation and the data source are listed in the second and third
  columns respectively.  Absolute magnitude (H) is listed along with
  the apparent magnitude (V) on the date of observation from
  \citet{bib.jpl09}.  We have determined the taxonomic type based on
  the SMASS \citep{bib.bus02a,bib.bus02c} classification system as
  described in the text.  The tight clustering in semi-major axis
  within each cluster is a consequence of their family membership.}
\label{t.Sub-My Asteroid Clusters}
\end{table}

Our spectroscopic reduction techniques for the new data followed
generally accepted practices using standard IRAF and IDL procedures
that are explained in detail in \citet{bib.wil08}.  Procedures applied
to some of the spectrographic data that may not be standard include:
\begin{itemize}
\item using 2-D arc lamp spectra to create a transform map to
  straighten the 2-D asteroid and analog spectra.
\item median-combining straightened dome flats into a column
  normalized master flat.  \ie the average of all pixels in any column
  (at the same wavelength) in the master dome flat was fixed at unity
  to correct for pixel-to-pixel variations in the quantum efficiency
  at all points on the CCD at the same wavelength.
\item creating a water band model by dividing solar analog spectra
  from the same star that show the greatest difference in water band
  amplitudes.  This allowed us to enhance the water bands for that
  solar analog.  Combining such cases from different stars produced
  the master water band spectrum that allowed better cancellation of
  atmospheric water absorption bands redward of 800 nm
  \citep{bib.bus02a}.
\item dividing the master water band spectrum into the asteroid
  spectrum and allowing the strength of the master to vary in order to
  minimize the distortion due to water bands. \citep{bib.bus02a}
\item binning the spectra into 10 nm bins in the manner of
  \citet{bib.bus02a} such that the realized resolution was in the
  range $44 \la R \la 92$ over the wavelengths spanning $440-920$~nm.
\end{itemize}

The spectral reduction procedures generally apply to UH2.2~m/SNIFS and
IRTF/SpeX spectra although some steps were performed automatically by
the SNIFS pipeline.  The IRTF/SpeX data reduction is facilitated by an
IDL based package called Spextool \citep{bib.cus04} which includes
preparation of calibration frames, processing and extraction of
spectra from science frames, wavelength calibration and flux
calibration.

We used the solar colors of \citet{bib.bla07} in our $PC_1$ and $PC_2$
measurement and absolute magnitude corrections for $u$ and $z$ bands
from \citet{bib.sds06b}.

To combine the spectral results with the photometric results described
below in a consistent manner we required a method to calculate $PC_1$
(eq.~\ref{eq.pc1}) for the spectra.  To do so we relied on the
relationship between spectral slope $s$ and $PC_1$ derived by
\citet{bib.wil08}.  They used a sample of 133 asteroids common to both
SMASS and the SDSS $3^{rd}$ Data Release to determine that
\begin{equation}
PC_1 = (0.87\pm0.02) \times (s/{\mu \rm{m}}^{-1}) + (0.082\pm0.012)
\label{eq.slopepc1}
\end{equation}

\noindent where the SMASS slope $s$ is determined over the wavelength
range from $0.44 - 0.92 \; \mu$m as the best fit line pivoting through
the normalization point at 0.55 \micron.

\begin{table}[!ht] \myfontsize
\begin{center}
\title{Table~\ref{t.ugriz}. Asteroid Pair Photometry}
\begin{tabular}{rlccccc}
& & \\
\tableline\tableline
 \multicolumn{2}{c}{$Asteroid$} & $u$ & $g$ & $r$ & $i$ & $z$ \\
\tableline
1986 & JN$_{1}$   & 18.40 $\pm$ 0.03 & 16.95 $\pm$ 0.03 & 16.46 $\pm$ 0.02 & 16.31 $\pm$ 0.01 & 16.24 $\pm$ 0.02 \\
2000 & WX$_{167}$ & 21.05 $\pm$ 0.08 & 19.65 $\pm$ 0.02 & 19.07 $\pm$ 0.02 & 18.92 $\pm$ 0.02 & 18.82 $\pm$ 0.04 \\
2001 & MD$_{30}$  & 18.95 $\pm$ 0.07 & 17.62 $\pm$ 0.02 & 17.10 $\pm$ 0.01 & 16.93 $\pm$ 0.01 & 16.89 $\pm$ 0.15 \\
{\bf 2000} & {\bf NZ$_{{\bf 10}}$}&{\bf 20.15 $\pm$ 0.06} &{\bf 18.50 $\pm$ 0.02} &{\bf 17.79 $\pm$ 0.02} & {\bf 17.61 $\pm$ 0.01} &{\bf 17.63 $\pm$ 0.03} \\
\smallskip
{\bf 2002} & {\bf AL$_{\bf 80}$}&{\bf 22.51 $\pm$ 0.30} &{\bf 20.66 $\pm$ 0.03} &{\bf 20.02 $\pm$ 0.02} &{\bf 19.83 $\pm$ 0.03} &{\bf 19.85 $\pm$ 0.07} \\
1999 & KF         & 21.21 $\pm$ 0.12 & 19.46 $\pm$ 0.02 & 18.77 $\pm$ 0.02 & 18.59 $\pm$ 0.02 & 18.58 $\pm$ 0.04 \\
2002 & GP$_{75}$  & 22.44 $\pm$ 0.27 & 21.01 $\pm$ 0.03 & 20.36 $\pm$ 0.03 & 20.27 $\pm$ 0.03 & 20.25 $\pm$ 0.11 \\
2006 & AL$_{54}$  & 21.97 $\pm$ 0.20 & 20.54 $\pm$ 0.03 & 19.88 $\pm$ 0.02 & 19.68 $\pm$ 0.02 & 19.66 $\pm$ 0.06 \\
\smallskip
1962 & RD         & 17.32 $\pm$ 0.05 & 15.54 $\pm$ 0.03 & 14.88 $\pm$ 0.03 & 14.60 $\pm$ 0.02 & 14.63 $\pm$ 0.02 \\
1997 & CT$_{16}$  & 21.14 $\pm$ 0.09 & 19.37 $\pm$ 0.02 & 18.59 $\pm$ 0.02 & 18.43 $\pm$ 0.01 & 18.44 $\pm$ 0.04 \\
2000 & RV$_{55}$  & 22.07 $\pm$ 0.27 & 20.31 $\pm$ 0.03 & 19.60 $\pm$ 0.02 & 19.36 $\pm$ 0.03 & 19.41 $\pm$ 0.05 \\
2004 & RJ$_{294}$ & 23.27 $\pm$ 0.54 & 21.77 $\pm$ 0.06 & 21.09 $\pm$ 0.05 & 20.93 $\pm$ 0.05 & 21.00 $\pm$ 0.19 \\
\smallskip
2003 & SC$_{7}$   & 22.21 $\pm$ 0.18 & 20.60 $\pm$ 0.02 & 20.03 $\pm$ 0.02 & 19.77 $\pm$ 0.03 & 20.15 $\pm$ 0.11 \\
2000 & GQ$_{113}$ & 20.52 $\pm$ 0.05 & 18.89 $\pm$ 0.03 & 18.27 $\pm$ 0.02 & 18.07 $\pm$ 0.02 & 18.24 $\pm$ 0.02 \\
1983 & WM         & 19.59 $\pm$ 0.03 & 17.78 $\pm$ 0.01 & 17.07 $\pm$ 0.01 & 16.87 $\pm$ 0.01 & 17.10 $\pm$ 0.02 \\
2003 & YK$_{39}$  & 22.53 $\pm$ 0.31 & 20.56 $\pm$ 0.03 & 19.83 $\pm$ 0.03 & 19.77 $\pm$ 0.02 & 19.81 $\pm$ 0.07 \\
\smallskip
1999 & TE$_{221}$ & 20.71 $\pm$ 0.08 & 19.21 $\pm$ 0.02 & 18.66 $\pm$ 0.02 & 18.46 $\pm$ 0.02 & 18.75 $\pm$ 0.04 \\
2000 & LU$_{15}$  & 21.50 $\pm$ 0.11 & 19.76 $\pm$ 0.02 & 18.99 $\pm$ 0.02 & 18.88 $\pm$ 0.02 & 19.39 $\pm$ 0.05 \\
2001 & XH$_{209}$ & 23.30 $\pm$ 0.56 & 20.58 $\pm$ 0.04 & 19.82 $\pm$ 0.02 & 19.57 $\pm$ 0.04 & 19.60 $\pm$ 0.10 \\
\tableline\tableline
\end{tabular}
\end{center}
\caption\small{$ugriz$ photometry for 19 asteroid pair members from
SDSS DR7 MOC4.  The two objects shown in bold constitute the only
complete pair.}
\label{t.ugriz}
\end{table}

\begin{figure}[ht!]
\centerline{\includegraphics[width=4.5in,angle=90]{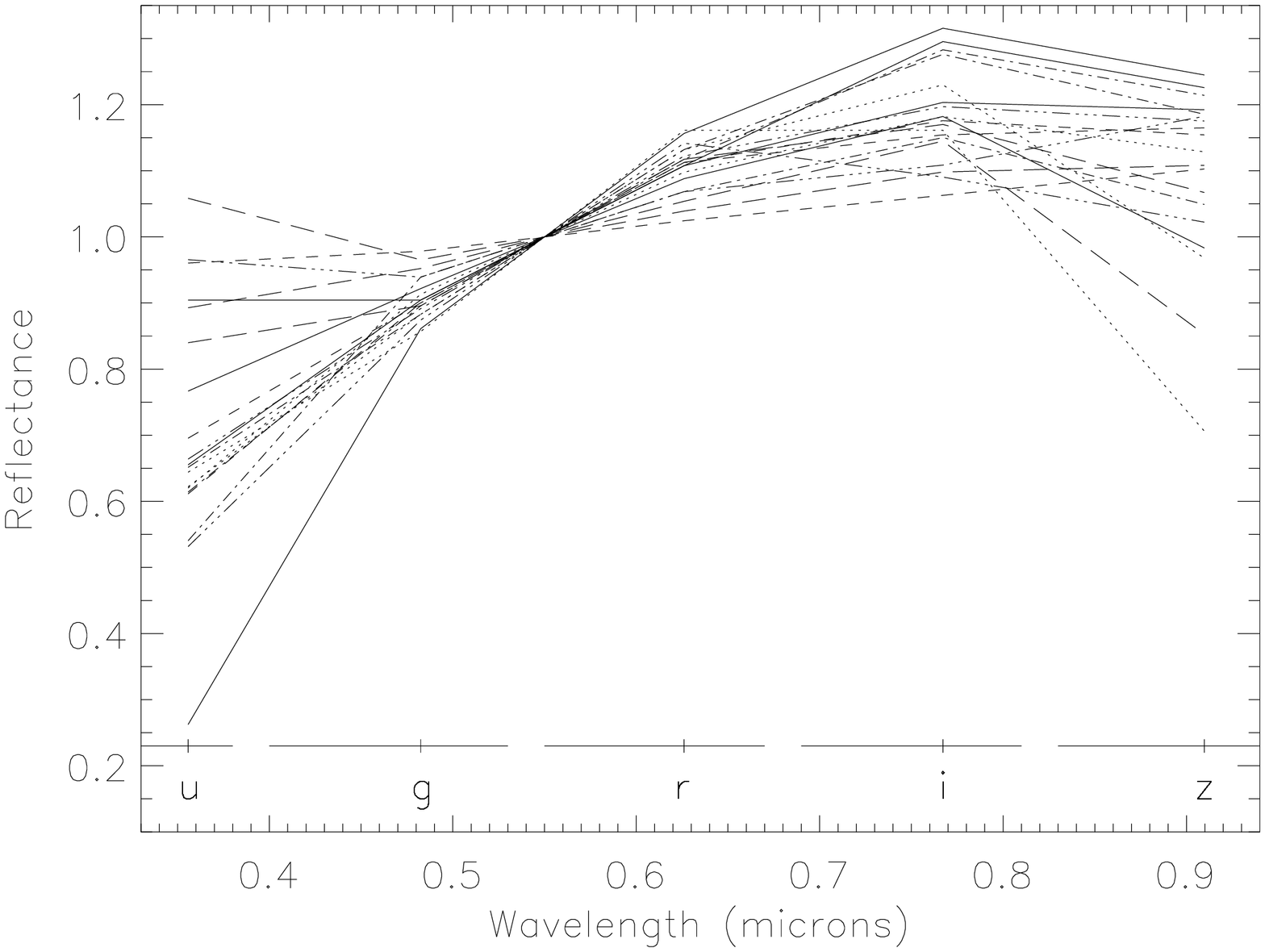}}
\caption\small{SDSS five-filter $ugriz$ photometry for 19 objects
common to the dynamically young pair members and also in the SDSSDR7
MOC 4.  The approximate band centers and widths for all five filters
are shown at the bottom of the Figure.  The SDSS filter measurements
have been connected by lines to simulate full spectra and are
normalized to 1.0 at 0.55 $\mu$m.}
\label{f.youngfluxes}
\end{figure}

Table \ref{t.ugriz} provides multiband photometry shown in Figure
\ref{f.youngfluxes} from the SDSS DR7 MOC4 for 19 members of 18 pairs
from the sub-set of non-family pairs \citep{bib.pra09b}.  Only one
pair had observations of both members available in the SDSS data set.

\section{Taxonomic type identification}
\label{s.classid}

While there has been recent and interesting progress in automated
taxonomic classification from asteroid photometry or
spectrometry\citep[\eg][]{bib.mar09b,bib.mis08,bib.bus00,bib.how94}
the process is still as much an art as science.  In an attempt to be
consistent (if not rigorous) in our taxonomic classifications we
developed and employed three closely related methods: 1) taxonomic
fitting 2) principal color component space location and 3) visual
assessment.  Our classifications with the three methods
were correlated with each other and with accepted taxonomic
classifications for known objects but were not identical.

The first method selects the best taxonomic `fit' for an object's
photometry to each SMASS spectral type $j$ according to the spectral
difference
\begin{equation}
\Delta_j^2 = \sum_{i=ugriz} \frac{(A_i -
S_{ij})^2}{(\sigma_{A_i}^2 + \sigma_{S_{ij}}^2)}
\label{eq.smassRank}
\end{equation}
\noindent where $A_i$ and $S_{ij}$ are the asteroid and SMASS
\citep{bib.bus02b} standard mean flux in band $i$ respectively and
$\sigma_{A_i}$, $\sigma_{S_{ij}}$ are the errors on the asteroid and
SMASS magnitudes in the same band.  The errors on the mean SMASS
spectra range from 0.01 - 0.06 and average about 0.03.

One complication with employing this technique is that the $u$ band
center is at 3557~\AA\ while SMASS spectra extend only over the range
4400-9200~\AA.  Therefore, to compare SMASS spectra to the SDSS filter
bands in eq.~\ref{eq.smassRank} (and in the other two methods
described below) we performed a linear extrapolation of the SMASS
spectra to the $u$ band central wavelength from the $g$ and $r$
band centers.  We tested several other methods of extrapolating to the
$u$ band including a quadratic extrapolation to $u$ using $g$, $r$
and $i$ bands or a quadratic extrapolation using all four of $g$,
$r$, $i$ and $z$ bands.  We also tested the result when we simply
ignored the $u$ band.  None of these techniques gave type
identifications as good as the simple linear extrapolation from $g$
and $r$ even though the linear approximation glosses over a possible
inflection point at 0.42 $\mu$m \citep{bib.ish07}.

\begin{figure}[!ht]\small
\centerline{\includegraphics[width=5.0in,angle=90]{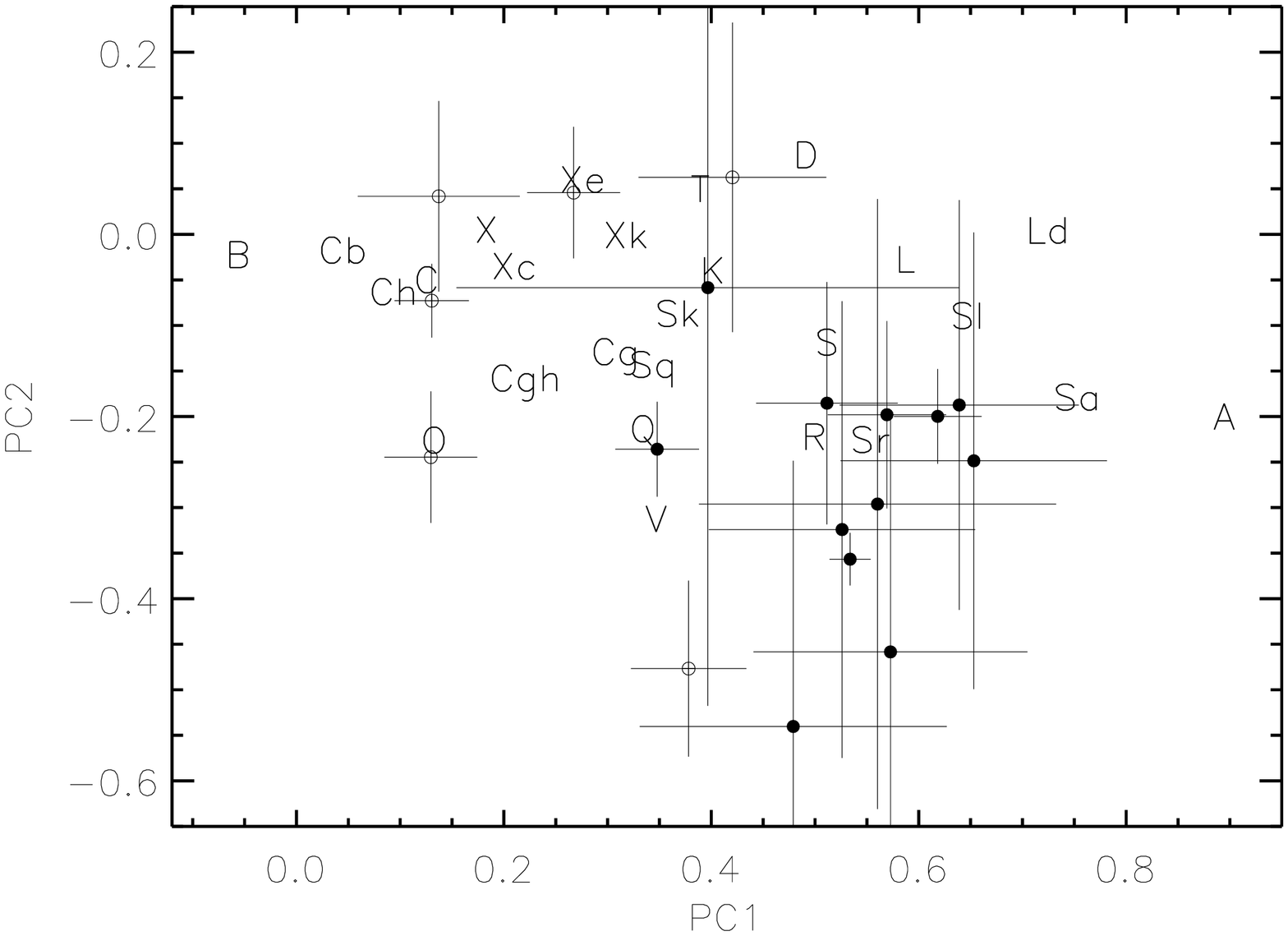}}
\caption\small{$PC_2$ vs. $PC_1$ for eighteen pair members having SDSS
  $ugriz$ photometry are shown as circles.  Filled/open circles lie
  inside/outside the region generously defining the S-complex.  Mean
  locations for SMASS classes are indicated by their taxonomic
  identification code.  We assign Q-class objects to the S-complex as
  discussed in the text.}
\label{f.pairspc1pc2}
\end{figure}

The second method selects the taxonomic type with the smallest
distance in principal component color space ($PC_1$,$PC_2$) between
the mean for each SMASS type $j$ and the object of interest as
illustrated in Figure~\ref{f.pairspc1pc2}.  We parametrize the
distance $R_j$ simply as
\begin{equation}
R_j^2 = \Delta PC_{1j}^2 + \Delta PC_{2j}^2
\label{eq.PCRank}
\end{equation}
\noindent where $ \Delta PC_{1j}$ and $\Delta PC_{2j}$ are the
difference in $PC_1$ and $PC_2$ between the object and the SMASS class
$j$ band average.

Finally, the third method was a traditional visual assessment of the
asteroid's spectrum.  This time-tested method incorporates elements of
both methods discussed above, takes into consideration the
extrapolation to $u$ band and takes advantage of human perception for
overall shape matching.  In this method, more weight was given to the
degree of shape matching than to minimizing band center differences.

In the end, we found that the top two candidate SMASS subclasses from
each of the three methods were consistent and that visual assessment
provided the the most reliable taxonomic classification method.

\section{Results and discussion}

\subsection{Sub-My cluster member taxonomy}
\label{ss.submyrClusterTaxonomy}

\begin{figure}[!ht]\small
\centerline{\includegraphics[width=5in,angle=90]{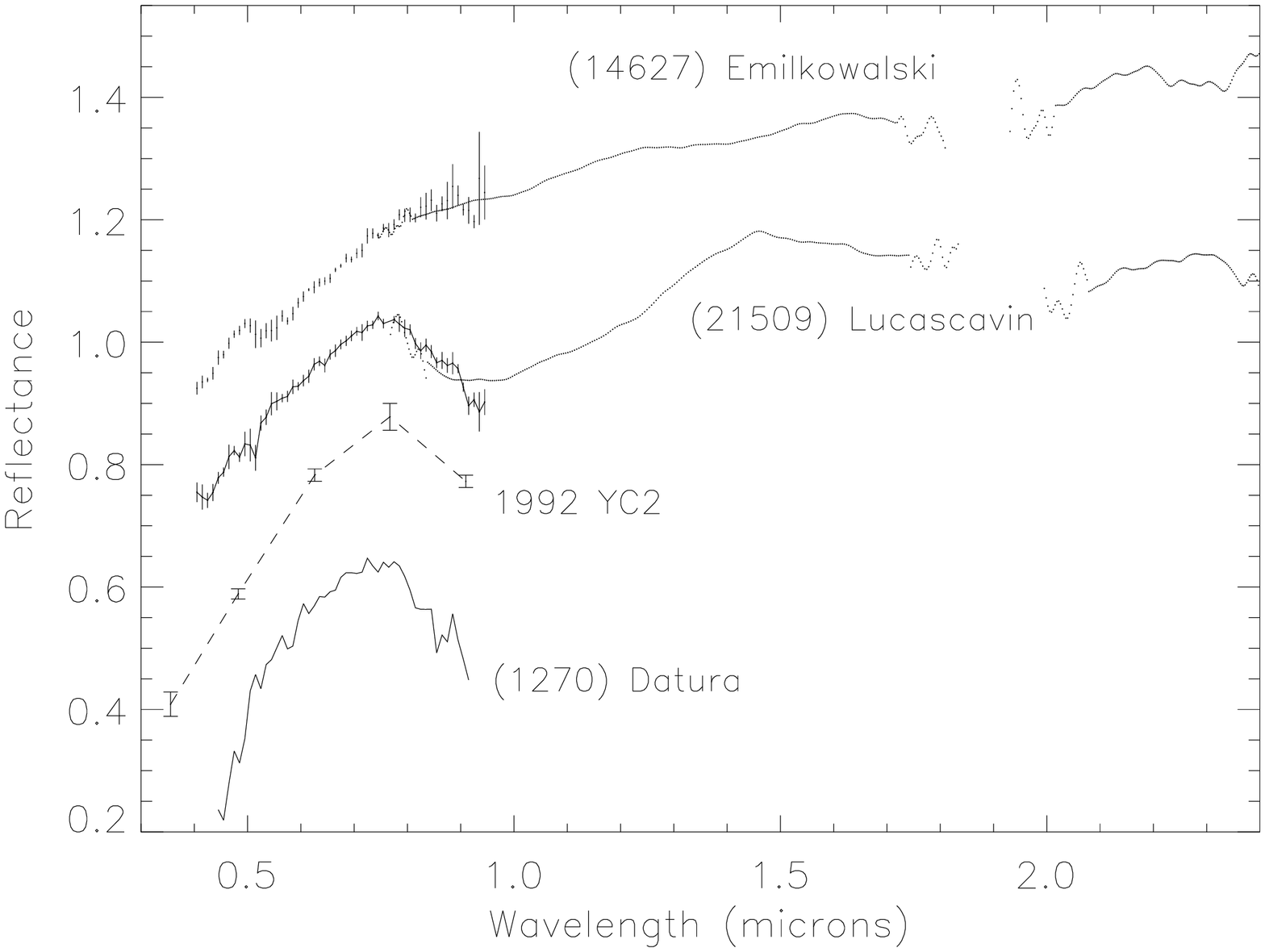}}
\caption\small{Visible and near IR spectra of (14627)~Emilkowalski and
  (21509)~Lucascavin obtained with UH2.2~m/SNIFS and IRTF/SpeX,
  (1270)~Datura visible spectrum obtained with UH2.2~m/SNIFS, and SDSS
  (16598) 1992 YC2 photometry.  The spectrum of (14627)~Emilkowalski
  is normalized to 1.0 at 0.55~$\mu$m and the others are offset
  vertically.  The gap at $\sim 1.9 \mu$m results from removing a sky
  absorption band.  The (1270)~Datura visible spectrum,
  (14627)~Emilkowalski and (21509)~Lucascavin IR spectra are all
  smoothed fits to the data (causing a spurious mismatch to the
  visible spectrum), the others are binned.}
\label{f.emillucdatvisir}
\end{figure}

We observed the brightest member of each of the four sub-My clusters
to identify their taxonomic type.  The spectra or multiband photometry
for each object are shown in Figure~\ref{f.emillucdatvisir}. (1270)
Datura, (16598) 1992 YC2 and (21509)~Lucascavin all show classic
S-complex characteristics in the visible --- a 0.75~$\mu$m peak and a
1.0~$\mu$m absorption band.  Datura also shows an inflection near 0.55
$\mu$m indicating a fresher surface relative to older S asteroids with
smoother spectra \citep{bib.ish07, bib.hir06}.

(21509)~Lucascavin also shows the 2.0~$\mu$m band typical of pyroxene.
Using the techniques described in \S\ref{s.classid} we identify
(1270)~Datura as an Sk, close to \citet{bib.mot08}'s identification of
Sl, (21509)~Lucascavin also as an Sk, and (16598) 1992 YC2 as member
of the S-complex.  Our visible and IR spectra of (14627)~Emilkowalski
does not show the 1.0 and 2.0~$\mu$m absorption bands typical of the
S-complex and we classify it as T type.  As our space weathering model
only applies to asteroids within the S-complex we ignore
(14627)~Emilkowalski for the remainder of this work.

It is unsurprising that (1270)~Datura and (21509)~Lucascavin were
identified in the S-complex as both clusters are located in the inner
main belt which is dominated by S-complex asteroids.  Similarly,
(14627)~Emilkowalski and (16598) 1992 YC2 are located in the middle of the
main belt where X and S-complex types are common.

Having identified three of the sub-My clusters within the S-complex
for which we intend to examine the effects of space weathering and
gardening we obtained data for eight of the cluster members as
provided in Table~\ref{t.Sub-My-cluster-pc1}.

\begin{table}[!ht]\small
\begin{center}
\title{Table~\ref{t.Sub-My-cluster-pc1}. S-complex sub-My cluster
members's derived spectral and age data}
\begin{tabular}{cccccc}
& & \\
\tableline\tableline
Cluster    & Asteroid   & Source   & Slope                & $PC_1$ & Age \\
           &            &          & (Reflectance/$\mu$m) &          & (kyr) \\
\tableline
           &(1270) Datura& MDN     & $0.355$              & $0.391$  & $530 \pm 20$ \\
           &(1270) Datura&this work& $0.382$              & $0.414$  & " \\
Datura     &(203370) 2001 WY$_{35}$& MDN&$-0.155$         &$-0.053$  & " \\
           & (60151) 1999 UZ$_6$   & MDN&$0.606$          & $0.609$  & " \\
           & (90265) 2003 CL$_5$   & MDN&$0.205$          & $0.260$  & " \\
\tableline\tableline
           &            &          &                      mean & $0.305 \pm 0.278$ \\
           &            &          &        \\
\tableline
           & (21509) Lucascavin&this work& $0.476$        & $0.496$  & $300-800$ \\
Lucascavin & (209570) 2004 XL$_{40}$&MDN                  & $0.314$  & $0.355$  & " \\
           & (180255) 2003 VM$_{9}$& this work            & $0.327$  & $0.366$  & " \\
\tableline\tableline
           &            &          &                 mean & $0.406 \pm 0.078$ \\
           &            &          &        \\
\tableline
1992 YC2   & (16598) 1992 YC2&MDN  & $0.083 \pm 0.02$     & $0.155 \pm 0.027$ & $135-220$ \\
\tableline\tableline
           &            &          &        \\
           &            &          & Sample Mean & $0.36 \pm 0.07$ & $511 \pm 10$ \\
\tableline\tableline
\end{tabular}
\end{center}
\caption\small{Derived color data and age estimates
  \citep{bib.nes06a,bib.nes06b} for members of three S-complex sub-My
  clusters.  $PC_1$ was calculated using eq.~\ref{eq.slopepc1}.
  Errors on the slope and $PC_1$ are not included as the error on the
  family's color is dominated by the distribution of $PC_1$ values
  within a cluster.  The errors on cluster means are standard
  deviations except for (16598) 1992 YC2 for which the measurement
  error is provided since it was the only object observed in the
  cluster.  The Sample Mean includes all eight cluster members with
  the error on the mean.}
\label{t.Sub-My-cluster-pc1}
\end{table}

\subsection{Taxonomy and orbit distribution of asteroid pair members}
\label{ss.pairMemberTaxonomy}

\begin{table}[!ht] \myfontsize 
\begin{center}
\title{Table~\ref{t.Asteroid Pairs}. Asteroid Pairs}
\begin{tabular}{rlcccrlccccc}
& & \\
\tableline\tableline
\multicolumn{2}{c}{$Ast_1$}   & SMASS  &Tholen&Quality& \multicolumn{2}{c}{$Ast_2$}& a      & e        & i       & $H_1$ & $H_2$ \\
   &        &        &      &     &     &       & (AU)    &          &(degrees)&         &         \\
\tableline
1986& JN$_{1}$   & X,  Xe & EMP, EMP & good & 2001& XO$_{105}$&$1.946$&$0.0601$& $23.710$& $13.5$ & $17.4$ \\
2000& WX$_{167}$ & Xe, T  & EMP, T   & fair & 2007& UV    & $1.909$ & $0.0613$ & $23.096$& $16.2$ & $17.1$ \\
2001& MD$_{30}$  & Xe, X  & EMP, EMP & fair & 2004& TV$_{14}$&$1.938$&$0.0886$ & $19.987$& $14.9$ & $17.2$ \\
{\bf 2000} & {\bf NZ}$_{{\bf 10}}$ & {\bf L, Sl} & {\bf S, S} & {\bf good} & {\bf 2002} & {\bf AL$_{{\bf 80}}$} & {\bf 2.287} & {\bf 0.1801} & {\bf 4.097} & {\bf 14.1} & {\bf 16.2} \\
\smallskip
{\bf 2002} & {\bf AL}$_{{\bf 80}}$ & {\bf Sl, S} & {\bf S, S} & {\bf good} & {\bf 2000} & {\bf NZ$_{{\bf 10}}$} & {\bf '' }   & {\bf '' }    & {\bf '' }   & {\bf 16.2} & {\bf 14.1} \\
1999& KF    & L, Sl & S, S     & good & 2008& GR$_{90}$   & $2.327$ & $0.2339$ & $1.777$ & $15.0$ & $17.2$ \\
2002& GP$_{75}$& L, S  & S, S  & good & 2001& UR$_{224}$  & $2.340$ & $0.1727$ & $3.865$ & $15.7$ & $17.2$ \\
2006& AL$_{54}$& L, Sl & S, S  & good & 2000& CR$_{49}$   & $2.272$ & $0.1763$ & $4.591$ & $16.8$ & $14.3$ \\
\smallskip
1962& RD    & Sl, Ld & S, S    & good & 1999& RP$_{27}$   & $2.198$ & $0.1775$ & $1.129$ & $13.1$ & $15.3$ \\
1997& CT$_{16}$  & Sl, Sa & S, S     & good & 2002& RZ$_{46}$& $2.186$ & $0.1672$ & $4.599$ & $15.4$ & $16.4$ \\
2000& RV$_{55}$  & Sl, Sa & S, S     & good & 2006& TE$_{23}$& $2.657$ & $0.1026$ & $2.245$ & $14.9$ & $16.8$ \\
2004& RJ$_{294}$ & S,  Sr & S, S     & good & 2004& GH$_{33}$& $2.268$ & $0.0981$ & $4.238$ & $18.2$ & $16.7$ \\
\smallskip
2003& SC$_{7}$   & Sk, K  & S, S   & good & 1998& RB$_{75}$  & $2.264$ & $0.1114$ & $7.263$ & $16.6$ & $14.6$ \\
2000& GQ$_{113}$ & Sq, Sk & S, S   & good & 2002& TO$_{134}$ & $2.324$ & $0.1319$ & $5.515$ & $14.4$ & $16.3$ \\
1983& WM    & Sr, Sa & S, S   & good & 1999& RC$_{118}$ & $2.320$ & $0.0790$ & $5.726$ & $13.7$ & $14.6$ \\
2003& YK$_{39}$  & Sr, Q  & S, Q   & good & 1998& FL$_{116}$ & $2.187$ & $0.0845$ & $3.736$ & $18.3$ & $15.0$ \\
\smallskip
1999& TE$_{221}$ & Q,  V  & Q, V   & fair & 2001& HZ$_{32}$  & $2.308$ & $0.1540$ & $5.642$ & $16.5$ & $15.0$ \\
2000& LU$_{15}$  & V,  Q  & V, Q   & good & 1992& WJ$_{35}$  & $2.313$ & $0.0701$ & $5.742$ & $16.1$ & $13.7$ \\
2001& XH$_{209}$ & A,  Sa & A, S   & good & 2004& PH         & $2.401$ & $0.2150$ & $3.638$ & $15.6$ & $16.4$ \\
\tableline\tableline
\end{tabular}
\end{center}
\caption{Asteroid pair member data for objects appearing in SDSS DR7
  MOC4.  We determined SMASS taxonomy for $Ast_1$ using $ugriz$
  photometry from SDSS DR7 MOC4 as described in the text.  The first
  and second ranked SMASS classes (\S \ref{s.classid}) are provided
  along with the corresponding Tholen classes and an assessment of
  that match's quality (degree of identification certainty).
  Semi-major axis (a), eccentricity (e), inclination (i) and absolute
  magnitudes ($H_1$,$H_2$) of both asteroids are from
  \citet{bib.vok08,bib.pra09b}.  Members shown in bold constitute the
  only complete pair.}
\label{t.Asteroid Pairs}
\end{table}

None of the members of 36 non-family asteroid pairs
(\citet{bib.pra09b, bib.vok08}, \S\ref{s.intro} describes paired
asteroids' discovery based on similar orbits.) are available in either
the SMASSI or SMASSII \citep{bib.bus02b} spectra databases or the
Eight Color Asteroid Survey (ECAS) \citep{bib.tho89}.  This is
unsurprising considering that the pair members are considerably
smaller than the typical asteroid in those surveys.  However, the 19
pair members identified in Table~\ref{t.Asteroid Pairs} were found in
the SDSS MOC4 \citep{bib.par08} from which we obtained the five-filter
solar-corrected $ugriz$ photometry in Table~\ref{t.ugriz}.

Most of the pair members are located in the inner main belt in a
region dominated by S-complex asteroids.  Their SDSS photometry
indicate that they belong to various taxonomic types typical of the
inner belt including L, S and V classes.  Our formal identification of
the pair member's taxonomy using the methods described in
\S\ref{s.classid} are provided in Table~\ref{t.Asteroid Pairs} which
shows that we have identified 14 of the 19 pair members with the
S-complex in which we also include L-class and Q-class.  We will
examine the colors of these asteroids in the context of our space
weathering model later in this section.  We also identified three
X-class, one V-class, and one A-class asteroid.

The taxonomic variety of the pair members is also represented in
Figures \ref{f.pairspc1pc2} and \ref{f.orbitaldistn} which shows that
their $PC_1$ color distribution is narrower than the full span of
SMASS classes while the distribution of their $PC_2$ values extends
beyond the SMASS class range.  Several pair members have $PC_2<-0.4$.
Since $PC_2$ corresponds roughly to a spectrum's curvature this
indicates an unusually convex spectrum.  A couple pair members lie
close to the V-class region while only one lies in the C-complex.  We
believe that the dearth of C-complex objects is an observational
artifact because pair members tend to be small and would be difficult
to detect with the low albedo of C-complex members in the outer belt.

The assignation of 1999~TE$_{221}$ to the Q class is important to our
space weathering analysis since it has been suggested
\citep{bib.bin04} that Q-class objects are actually very young,
essentially unweathered, S-complex asteroids.  Thus, we assume that it
is a particularly young member of the S-complex with a deep $1 \mu$m
band as has been predicted for young S-complex asteroids.  However, in
osculating element space it is located close to 2000~LU$_{15}$,
another member of one of our asteroid pairs from table \ref{t.Asteroid
Pairs} that we assigned to the V-class.  Both the asteroids lie close
to the edge of the Vesta family region as shown in
fig. \ref{f.pairspc1pc2} (again, in osculating elements). This opens
the possibility that 1999~TE$_{221}$ could be a Vestoid with a
slightly shallower $1 \mu$m absorption band.

To confirm our Q-class identification for 1999~TE$_{221}$ and as an
additional check on our type-identifications we examined whether the
pair members are of taxonomic types typical of their orbit element
phase space region.  To do so we identified each pair member's five
nearest orbit element neighbors (using the $D$-criterion of
\citet{bib.nes05}) in the set of 1175 objects from SMASS that also
have osculating orbital elements in \citet{bib.ast08}.  We found that
17 out of 19 pair members match their nearest neighbor's complex
suggesting that our identification methods identify the right complex
$\sim 90$\% of the time and that the pair members are representative
of the composition of the main belt region in which they are located.

As mentioned earlier, this test was particularly important for the
cases of 1999~TE$_{221}$ and 2000~LU$_{15}$ as both lie on the
periphery of the Vesta family region.  We identify the $1^{st}/2^{nd}$
most likely types for these two objects as Q/V and V/Q respectively.
1999~TE$_{221}$'s five nearest neighbors include four in the S-complex
with one being a Sq and none in the V-class.  On the other hand,
2000~LU$_{15}$ has three V-class neighbors.  This supports our ability
to reliably distinguish Q from V.  Remember that we place the Q-class
within the S-complex and, since 1999~TE$_{221}$ is by far the bluest
member of the S-complex pair members, its inclusion in our analysis
could have a substantial impact on the mean $PC_1$ of the pair members
and on our measurement of the space weathering rate of S-complex
asteroids.  The effect of including or excluding 1999~TE$_{221}$ in
our analysis is described later.

\begin{figure}[!ht]\small
\centerline{\includegraphics[width=4.5 in,angle=90]{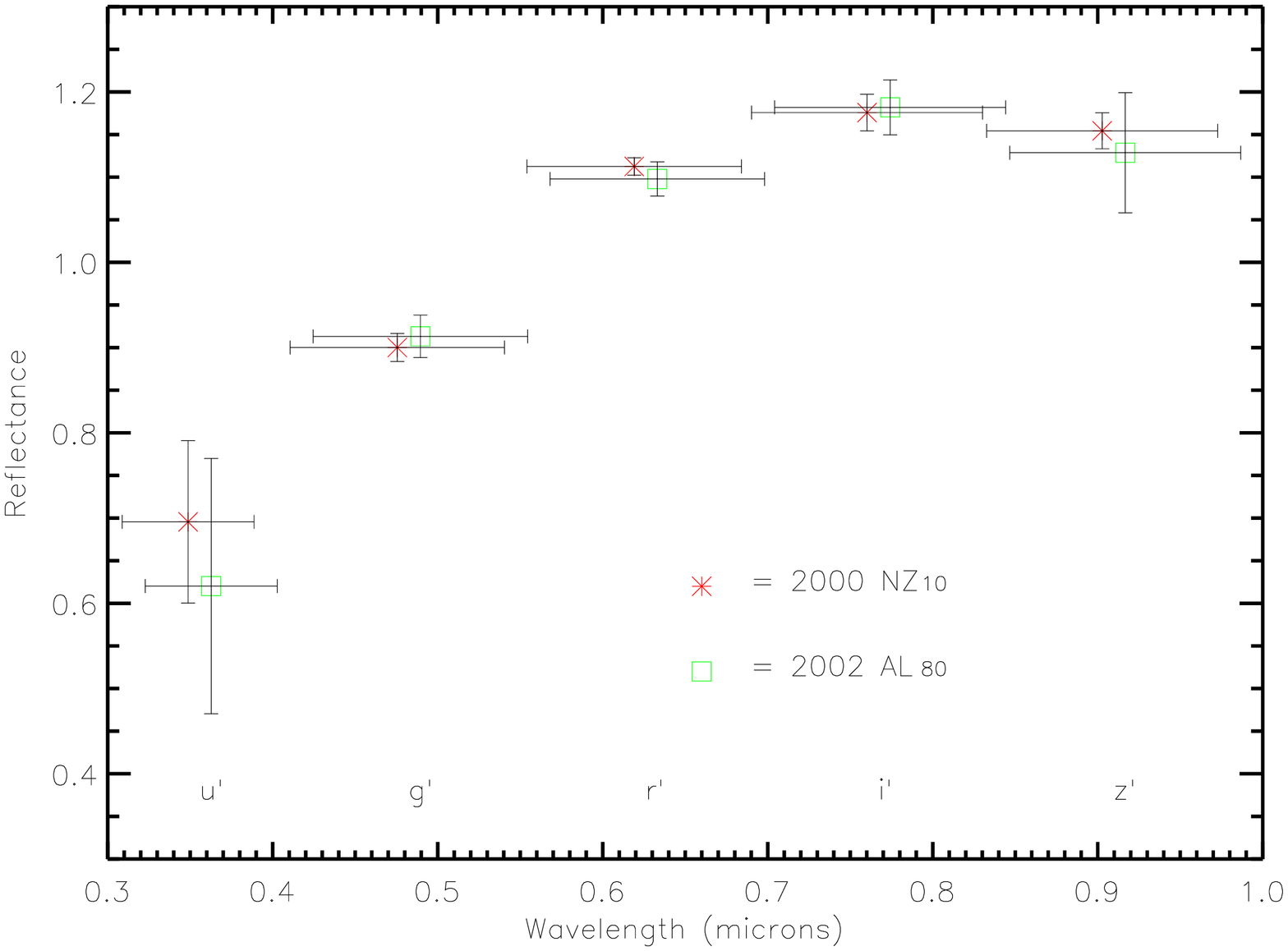}}
\caption\small{SDSS filter photometry for both members of the only
  complete dynamical pair in the MOC4.  The central wavelength for
  each data point corresponds to the band centers for the $ugriz$
  filters except for a small horizontal offset for clarity while their
  width represents the band pass.  The data is normalized such that a
  straight line between the $g$ and $r$ data points passes through
  unity at 0.55~$\mu$m.}
\label{f.matchpair}
\end{figure}

\begin{figure}[!h]\small
\centerline{\includegraphics[width=5.0in,angle=90]{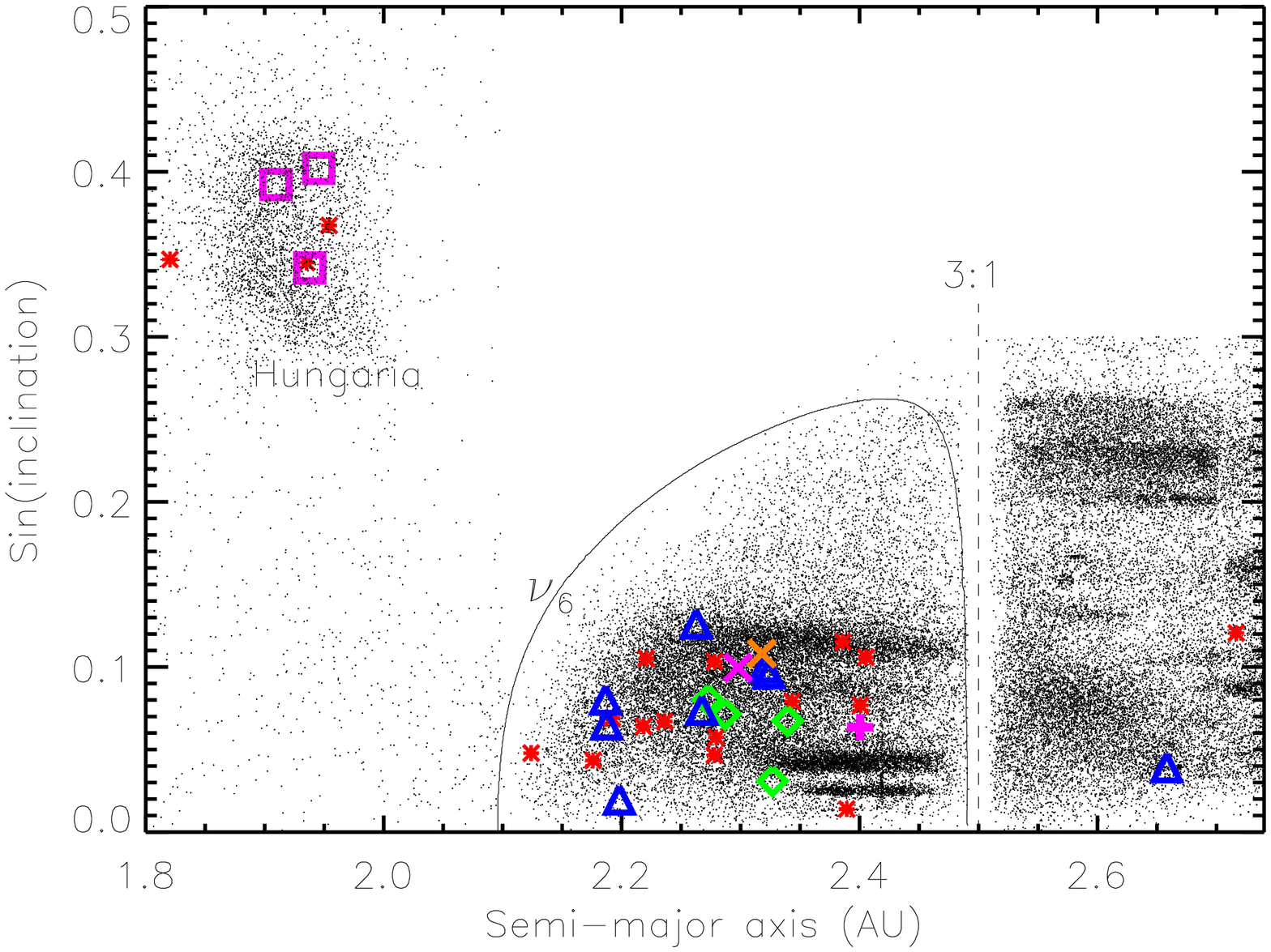}}
\caption\small{Osculating $\sin$(inclination) versus semi-major axis
  for 36 pair members (large colored points) superimposed on the
  proper element distribution for main belt asteroids identified in
  the SDSS MOC4 (black dots).  For semi-major axis $<2.1$~AU we show
  osculating elements for main belt asteroids from Astorb
  \citep{bib.orb08}.  The 18 distinct pair members with SDSS
  photometry were identified as the following types: violet squares
  are X-complex, blue triangles are S-complex, green diamonds are
  L-complex, the orange $\times$ is V-class, the fuchsia $\times$ is
  Q-class, and the fuchsia $+$ is A-class.  The 18 red asterisks
  represent pairs for which neither member is present in the SDSS
  MOC4.  The $\nu_6$ and 3:1 resonances are shown for orientation
  along with the Hungaria family region.}
\label{f.orbitaldistn}
\end{figure}

Our interest in and utilization of the asteroid pairs for the purpose
of measuring young asteroid surface ages assumes that the pair members
are genetically related and fissioned by some as yet undefined process
$<0.5$ My ago \citep{bib.vok08}.  If the members of a pair are
genetically related asteroids then our expectation is that they will
display nearly identical spectra.  Only one complete pair
(2000~NZ$_{10}$ and 2002~AL$_{80}$, see Table \ref{t.Asteroid Pairs})
was identified among the 19 pair members available in the SDSS MOC4.
Figure~\ref{f.matchpair} shows that the colors of the two objects
match and therefore supports a genetic origin of the pair.  Using the
taxonomic identification methods of \S\ref{s.classid} we find that
2000 NZ$_{10}$ is SMASS L-class and 2002 AL$_{80}$ is Sl-class -
adjacent classes in $PC_2$ vs. $PC_1$ color space as shown in
Figure~\ref{f.pairspc1pc2}.  Thus, this one line of evidence suggests
that asteroid pairs are genetically related.

Having established the taxonomic composition of the asteroid pairs and
their likely genetic relationship we would like to examine their
taxonomic-orbit distribution --- does the pair member taxonomy match
that of their neighbors in orbit element space?  The answer to this
question could shed light on the relative internal strengths of the
different types or provide information on the mechanism for asteroid
pair creation.  \ie if C-complex asteroids split into pairs more
frequently it could imply that they are weaker than other types or
that the pair formation mechanism acts more efficiently on them.
Unfortunately, answering this question is beyond the scope of this
work and we leave it to the future.  Instead, we make a couple simple
observations on the pair's orbit element distribution.

The axis-inclination structure of the main belt and the pairs is shown
in Figure~\ref{f.orbitaldistn} revealing that the 36 pairs are
distributed in two clumps; a high inclination clump inside 2.0~AU
within the Hungaria family region, a group dynamically protected from
perturbations by Mars via their high inclination, and a clump on the
inner edge of the main belt with 2.1~AU$\lae a \lae$~2.4~AU. There are
also two outliers in the middle belt with semi-major axes in the range
2.65~AU$<a<$~2.75~AU.  That most of the pairs are located in the
innermost main belt is almost certainly an observational selection
effect --- asteroid pairs are composed of small asteroids that are
only visible when they are located close to the Earth \ie in the inner
main belt.

The Hungaria clump includes six pairs of which three have SDSS
photometry that we identify as (SMASS) X-complex asteroids ---
consistent with \citet{bib.gra82}'s claim that roughly 70\% of the
asteroids in the Hungaria region are Tholen E or R class (the Tholen E
class is contained within the SMASS X-complex).  Thus, our
identification of three SDSS X-complex members in the region is
unsurprising and provides further support for our taxonomic
classification techniques.

\subsection{Space weathering on sub-My clusters and asteroid pairs}
\label{ss.comparemodel}

\begin{figure}[!ht]\small
\centerline{\includegraphics[width=5in,angle=90]{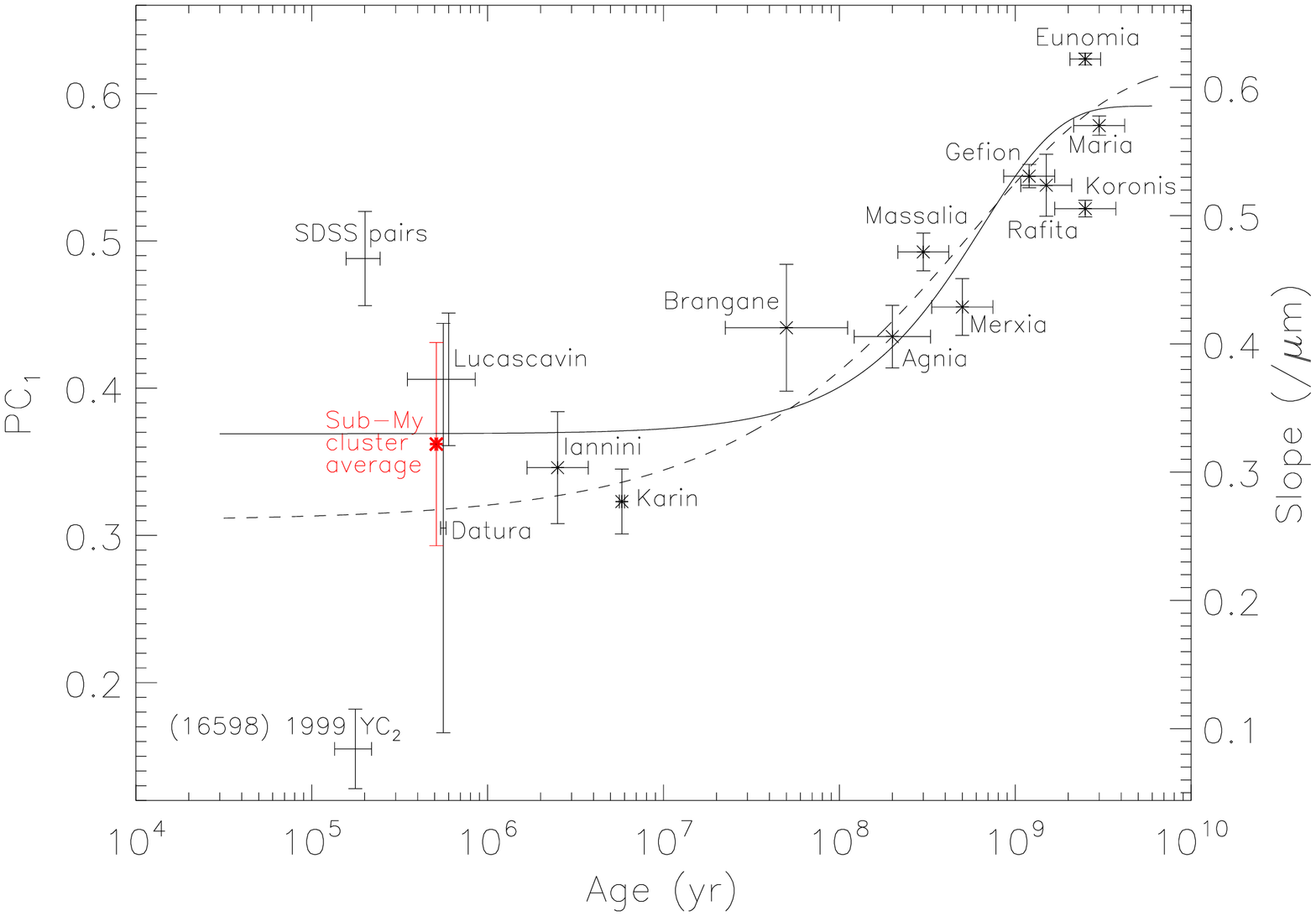}}
\caption\small{$PC_1$ color and dynamically determined ages for
  S-complex asteroid families adapted from \citet{bib.wil08}.  (The
  corresponding spectral slope is shown on the right.)  The dashed
  curve represents their space weathering model (eq.~\ref{eq.jedpc1})
  extrapolated to the sub-My region.  Three S-complex sub-My clusters
  ((1270)~Datura, (21509)~Lucascavin, (16598) 1992 YC2) are shown
  individually and with their mean value indicated by the red `Sub-My
  cluster average' data point.  The solid curve represents the dual
  weathering/gardening model fit to the family data including the
  sub-My cluster point.  The `SDSS pairs' point represents the average
  color of 12 unique S-complex sub-My pairs found in SDSS DR7 MOC4.
  Errors are standard errors on the mean except for (16598) 1992 YC2,
  a single object, for which we provide the measurement error.}
\label{f.spaceWeathering}
\end{figure}

Figure~\ref{f.spaceWeathering} combines our previous work
\citep{bib.wil08} with the new color-age data in this work for
S-complex asteroids in three sub-My clusters and eleven sub-My
asteroid pairs.  The mean $PC_1$ for the three sub-My clusters is
$0.36 \pm 0.07$ --- within 1$\sigma$ of \citet{bib.wil08}'s predicted
color of $PC_1 = 0.31 \pm 0.04$ for the clusters's mean age of $440
\pm 60$ kyr.  The good agreement with the prediction lends confidence
to the space weathering model which now agrees with the cluster
color-age data over five orders of magnitude in age in the decades
from $10^5 - 10^{10}$ years.

However, the weighted mean $PC_1 = 0.49\pm0.03$ color (the error is
weighted error on the mean) for the S-complex young pairs is over
$5\sigma$ redder than predicted by \citet{bib.wil08}.  Excluding the
Q-type pair member 1999~TE$_{221}$ discussed in
\S\ref{ss.pairMemberTaxonomy} increases the mean $PC_1$ to $0.54 \pm
0.04$ and the discrepancy to over $7\sigma$.  The very young asteroid
pairs clearly do not follow the space weathering function proposed by
\citet{bib.wil08}.

The disparity may be explained in a number of ways: 1) the asteroid
pairs do not represent a recent breakup of a parent body and are not
genetically related or 2) the asteroid pairs are the result of a
recent breakup but with only partial resurfacing which did not `reset'
the space weathering clock or 3) the space weathering model of
\citet{bib.wil08} is either too simplistic or wrong.  We examine each
of these scenarios in turn:
\begin{enumerate}
\item We consider it unlikely that the asteroid pairs are not
  genetically related for two reasons.  First, the pairs were specifically
  selected \citep{bib.vok08} because they are statistically likely to
  be related asteroids.  Furthermore, the colors of the pair for which
  both members exist in the SDSS MOC4 agree extremely well (see
  Fig. \ref{f.matchpair}).  While asteroids that inhabit the same
  region of the main belt often have similar colors the scale of
  agreement in both the orbit and colors argues persuasively for a
  genetic link between the pair members.
\item Some pair formation scenarios may not be as violent as the
  formation of large asteroid families through the catastrophic
  disruption of a parent body and may not reset the entire surface to
  zero age.  \citet{bib.vok08} cite three possible methods of forming
  pairs: catastrophic collision followed by fragment reaccumulation
  into binary orbits \citep[\eg][]{bib.dur04,bib.nes06c}, YORP induced
  rubble pile spin-up leading to calving of the secondary
  \citep[\eg][]{bib.wal08}, or YORP induced angular acceleration of
  contact binaries leading to their separation
  \citep[\eg][]{bib.mer02,bib.pra09a,bib.dur04}.  The first scenario
  will `reset' the age of the entire surface of both the primary and
  secondary.  The third mechanism might leave large portions of the
  surface undisturbed since the only portion that is necessarily
  exposed is the binary contact region.  It is unclear how much of the
  surface would be affected by the second scenario.  Therefore there
  is a clear distinction between the possible formation mechanisms of
  pairs and larger groupings.  Asteroid families and clusters with
  more than two members form only by the first catastrophic
  collisional scenario while pairs can also form by the two variations
  on YORP spin-up.  This yields at least one pair formation scenario
  that could leave some of the members' surface undisturbed.  In this
  case the globally averaged surface color would correspond to a
  misleading age somewhere between fresh surface and the age of the
  parent body's surface.  Taking the color of the pairs at face value
  and interpreting it in the framework of the \citet{bib.wil08} space
  weathering model indicates an average surface age of $\sim 440$~My
  --- over 1000$\times$ older than their dynamical age.  If we assume
  that the original parent body's surface was reddened to saturation
  prior to separation, and taking the dynamical ages of $<500$~kyr for
  the pairs at face value such that freshly exposed surface is
  essentially unweathered, then $\sim 64$\% of the surface must be
  disturbed in the pair separation process.  Considering that it is
  unlikely that the surface was fully weathered prior to separation
  allows us to set an upper limit on the fraction of disturbed surface
  at $\la 64$\%.  (Excluding the Q-type object 1999~TE$_{221}$ only
  changes the upper limit to $\la 48$\%.)  A gentle binary separation
  due to slow YORP spinup may be consistent with this scenario.
  Indeed, \citet{bib.pra09a} also provide evidence from reconstruction
  of the initial configuration of the 6070-54827 pair and rotation
  rate observations that the non-family pair formation process is a
  gentle event. If we envision a bi-lobed asteroid gradually
  accelerating in angular velocity and finally fissioning at the neck
  that joins the lobes then it seems reasonable that the portion of
  disturbed surface would be $\la64$\%.  In the binary/pair formation
  mechanism proposed by \citet{bib.wal08,bib.wal06} an asteroid's
  polar surface material migrates to the equator as the object's
  rotation rate increases and eventually flies into orbit around the
  parent body where it reaccumulates into a satellite.  The primary
  and satellite eventually separate and their orbits evolve
  dynamically.  Our impression is that this model would generate fresh
  (blue) surface on both the primary and satellite in conflict with
  our observation of reddish surfaces on the asteroid pairs.  However,
  it is not difficult to envision a slow migration process that allows
  material to weather on the primary's surface before being shed and
  accumulating into a secondary object.
\item It is possible that the space weathering model of
  \citet{bib.wil08} that built upon the earlier work of
  \citet{bib.nes05} and \citet{bib.jed04} is simply wrong; that the
  apparent change in color of S-complex asteroids with age is a
  statistical fluke or due to some other underlying effect (though
  obvious possibilities were considered in detail in the early works).
  However, \citet{bib.par08} confirm the weathering effect in an
  independent updated analysis of the SDSS DR7 MOC4 data.  We consider
  it more likely that the space weathering mechanism is more
  complicated than simply affecting the average spectral slope (or
  $PC_1$) as a function of time.  For instance, it is well known that
  space weathering affects not only the slope of the spectrum but also
  the depths of the 1~$\mu$m and ultraviolet absorption bands and the
  surface albedo.  We consider it not only possible but likely that
  these effects occur during space weathering at different rates.  The
  apparent redness of the young asteroid pair members relative to the
  expectation of the simplistic space weathering model could indicate
  that a `fast' space weathering process takes place in $\la 10^5$
  years.  However, this scenario requires a rather contrived sequence
  of events.  Consider the three color change processes: decreasing
  depth of the ultraviolet band shortward of 0.4~$\mu$m, decreasing
  depth of the 1~$\mu$m absorption band and continuum reddening
  between these two bands.  The first process is the only one that
  produces a bluer color.  Therefore, accounting for the anomalous
  redness of the pair members requires the unlikely scenario that one
  of the latter two processes dominate on short time scales which is
  then belatedly overwhelmed by the first process which is then
  finally dominated by the third.
\end{enumerate}
Given the disagreement between the asteroid pair color-age and the
\citet{bib.wil08} space weathering model, and considering our
enumerated arguments above, we continue with our analysis under the
assumption that we can ignore the colors of the asteroid pairs in this
new determination of the space weathering and gardening rates.

First, we fit all the S-complex color data including the sub-Myr
clusters but not the asteroid pairs to the `old' space weathering
function of eq. \ref{eq.jedpc1}.  Considering the good agreement
between the predicted \citep{bib.wil08} and observed colors of the
sub-Myr cluster members it is unsurprising that the fit including the
new data matches the previous fit in all four parameters to within
1-$\sigma$ with $PC_1(0)=0.34 \pm 0.02$, $\Delta PC_1=0.28 \pm 0.05$,
$\tau = 700 \pm 270$~My, $\alpha = 0.58 \pm 0.17$.  However, as we
observed in \citet{bib.wil08}, fitting the color-age data to the form
of eq. \ref{eq.jedpc1} suffers from multiple and wide minima in the
fit-parameter space.  Furthermore, the function does not explicitly
separate the weathering and gardening effects.

On the other hand, we found that fitting the same color-age data to
our new function that incorporates both space weathering and gardening
(eq. \ref{eq.garden}) is better behaved because the solution space
does not show multiple local $\chi^2$ minima.  The best fit (lowest
$\chi^2$) including the sub-My clusters (but, again, not the asteroid
pair colors) yields $PC_1(0)=0.37 \pm 0.01$, $\Delta PC_1=0.33 \pm
0.06$, $\tau_w = 960 \pm 160$~My, $\tau_g = 2000 \pm 290$~My as shown
by the solid curve in Figure~ \ref{f.spaceWeathering}.

Note that it is not correct to compare the new $\tau_w$ or $\tau_g$ to
the old $\tau$ value because of the complicating and non-physical use
of $\alpha$ in the generalizing exponent in the old functional form.
In essense, the `old' value was an effective weathering time that
combined the effects of both regolith gardening and weathering.  Since
the effective weathering time can be shown to be equivalent to the
time corresponding to the inflection point on the new weathering
function, Figure~\ref{f.spaceWeathering} shows that the two models are
in good agreement.

The old and new $\Delta PC_1$ are not strictly comparable either.
Formerly in the weathering only case the entire surface would
eventually reach the color $PC_1 = PC_0 + \Delta PC$.  The
weathering/gardening case will produce a lower equilibrium value with
$PC_1 < PC_0 + \Delta PC$.

The new space weathering-only time frame of $\sim$1~Gy is consistent
with the `slow weathering' measured in lab-based measurements as
discussed in the introductions. The `slow' school includes our result
of $\sim$1~Gy based on space observations of S-complex families,
\citet{bib.sas01}'s equivalent value of 700~My at 2.6~AU based on
pulsed-laser irradiated silicate pellets, and \citet{bib.pie00}'s
estimate of $100-800$ My for lunar surfaces based on crater counts and
spectra and color-terrain correlations on (243) Ida \citep{bib.vev96}.
But these slow space weathering results disagree with the `fast
weathering' school that includes lab-based ion irradiation results
from \citet{bib.ver09} ($\lae$~1~My) and \citet{bib.loe09}'s ($\sim$ 5
kyr), and \cite{bib.tak08}'s $<$450~kyr time scale based on a shallow
1 $\mu$m absorption band observed on (1270) Datura.  We do not have an
explanation for the discrepancy between the fast and slow weathering
results.

\subsection{Gardening Time}
\label{s.gdntime}

In \S\ref{s.gardenimpact} we calculated the gardening time for
asteroid regolith as a function of target diameter as shown in
Figure. \ref{f.gardentaulog}.  Over a broad range of asteroid sizes
the gardening time scale is $200-300$ My years.  This result is in
dramatic disagreement with a similar calculation by \citet{bib.mel09}
that was based on a collisional cratering model by \citet{bib.gil02}
and yielded a time scale of 2000 years for resurfacing Trojan
asteroids.  It is difficult to reconcile the two results that differ
by five orders of magnitude.  Two factors that we are aware of that
explain some of the discrepancy are 1) that they used a Trojan impact
probability double that of the main belt and 2) the slope of the
Trojan size-frequency distribution was assumed to be $-3$ whereas the
data we used had slope of $\sim -2.20$ appropriate to the main belt
\citep{bib.bot05b}.  But these two differences only account for a
fraction of the difference between our results.

Figure \ref{f.gardentaulog} also shows that our measured gardening
rate from the S-complex asteroid color-age relationship is an order of
magnitude different from our calculated rate based on impacts.  Is it
possible to reconcile this difference?

There is considerable uncertainty in the various terms involved in
calculating the impact gardening time from eq. \ref{eq.tauofR} but we
have done our best to select the best contemporary values in each
case.  While the gray region on the figure represents the formal
1-$\sigma$ error on the calculation based on the reported errors on
each input quantity there is considerable unreported systematic error
in the calculation associated with its sensitivity to the input
parameters and functions.  In particular, we examined the sensitivity
of the calculation to the:
\begin{itemize}
\item asteroid size frequency distribution, $N(D)$
\item specific shattering energy function, $Q_S^*$, that determines
  the largest non-disruptive impactor (we used the \citet{bib.mel97}
  function as a comparison),
\item impact speed and,
\item $D_{min}$, the smallest impactor that creates ejecta
\end{itemize}

In each case we varied the input parameter or function over a range of
of 2 to 4 in each direction.  In most cases the gardening rate for
large target objects is only slightly affected.  The rate for objects
$\sim$1~km diameter changed by a factor of about two or slightly more.
The two most important factors in determining the gardening rate are
1) $D_{min}$ and 2) the amount of area covered by crater ejecta.
Increasing $D_{min}$ or decreasing the ejecta area coverage both work
to increase the gardening timescale.  Leaving all other parameters at
their nominal values, increasing $D_{min}$ to $\sim$30~m is sufficient
to increase the gardening time to our measured value of $\sim$2~Gy.  A
similar result is obtained by decreasing the diameter of the ejecta
field by a factor of two (the affected area by a factor of four).

In a related observation, \citet{bib.cha05} and others note the
unexpected paucity of craters $<$200 m diameter on (433)~Eros.  Taken
at face value the observation would imply a deficit of impactors $\la
20$~m in diameter but \citet{bib.obr09} has shown that no such deficit
can exist because it would in turn generate a `wave' in the observed
size-frequency distribution for larger asteroids that is not observed.

Thus, we have identified two independent problems --- the mismatched
measured color-age and impact-calculated gardening times and the
crater deficiency on (433)~Eros --- that can both be resolved if
impactors $\la 20$ m do not leave a crater record.  While
\citet{bib.ric04,bib.ric05} have proposed that seismic shaking can
erase the small craters the mechanism can only explain the crater
shortage not the gardening time mismatch.  This is because any process
that increases the gardening rate (\ie seismic shaking) shortens the
calculated impact gardening time and worsens the mismatch with
measured color-age gardening rate. 

If we believe that there is no deficit of impactors of $\la 20$~m
diameter and that there is a deficit of craters and regolith gardening
caused by impactors in the same size range then it must be the case
that the impacts take place without creating craters in the `normal'
manner.  Perhaps the surfaces of small asteroids `absorb' the small
impactors in an inelastic collision without creating a crater an order
of magnitude larger or ejecting significant material.  Like throwing a
small stone into a bee's empty honeycomb.

\section{Summary}

We have combined data from five asteroids obtained on KeckII/ESI,
UH2.2/SNIFS and IRTF/SpeX, spectra of six asteroids from
\citet{bib.mot08} and SDSS archive photometry on 19 asteroid pair
members to investigate the earliest stages of space weathering on
asteroids aged less than one million years, including sub-My old
asteroid clusters and asteroid pairs.

\citet{bib.wil08} predicted that the color of these young S-complex
asteroid's surfaces would be essentially identical to freshly exposed
regolith with $PC_1 = 0.31 \pm 0.04$.  Our measured color data
for asteroid clusters with ages in the range $10^5 - 10^6$ years
agrees with their prediction with a mean $PC_1 = 0.36 \pm 0.07$.

For the first time, we extended the space weathering model to
explicitly include the effects of regolith gardening which restores
weathered S-complex asteroid surface to its original blue color.
Fitting the refined model to the former data along with the new sub-My
asteroid cluster data point gives separate characteristic times for
weathering and gardening of $\tau_w = 960 \pm 160$ My and $\tau_g =
2000 \pm 290$ My respectively.  The new results suggest that fresh
S-complex asteroids have a blue surface color with $PC_1(0)=0.37 \pm
0.01$ and would redden by $\Delta PC_1=0.33 \pm 0.06$ over a very long
period of time in the absence of gardening.  The presence of gardening
produces an ultimate equilibrium color of $PC_1(\infty)=0.59 \pm
0.06$.  The new weathering time scale and colors are consistent with
the results of \citet{bib.wil08}.

Our new data for asteroid pair members with ages $\lae 10^5$ years is
redder than predicted by more than $5\sigma$.  Assuming that the space
weathering model is correct, the discrepancy could be due to the pair
production formation mechanism --- if the asteroids's surfaces are not
fully recoated during a gentle separation then the surface age for
these pairs could be much older than the dynamical age since
separation.  This explanation requires that $\la 64\%$ of the
asteroids's surface is disturbed during the pair formation event
providing an interesting test in comparison to simulations of pair
formation.  An alternative explanation that there are `fast' and
`slow' weathering processes occuring on S-complex asteroid surfaces
does not seem plausible.

We independently calculated the gardening rate on main belt asteroids
from basic principles including the affects of the asteroid
size-frequency distribution, impact rates, crater and ejecta
formation.  The calculated gardening time scale of $\sim 270$ My for a
$D \sim$3 km asteroid typical of those in our sample is in stark
contrast to the color-age measured value of $\sim 2000$ My.  Two
scenarios that can reconcile the calculated and measured gardening
rates are 1) impacting asteroids of $\la 20$~m diameter do not produce
craters in the standard manner and 2) the ejecta field is much smaller
than standard models would suggest.  Perhaps the smaller asteroids are
absorbed in inelastic collisions with a `honeycomb'-like surface.
This mechanism also provides an explanation for the paucity of small
craters on (433)~Eros.

\section{Acknowledgments}
This work was supported under NSF grant AST04-07134. The new data
presented herein were obtained with Keck II/ESI, UH2.2m/SNIFS and
IRTF/SpeX.  Greg Aldering and the entire SNIFS team helped train us on
their instrument.  Yannick Copin assisted in SNIFS data reduction for
the early sub-My results.  Bill Bottke provided simulated impact rate
data and helpful discussion about cratering.  Some of the data
presented herein were obtained at the W.M. Keck Observatory, which is
operated as a scientific partnership among the California Institute of
Technology, the University of California and the National Aeronautics
and Spcae Administration.  The Observatory was made possible by the
generous financial support of the W.M. Keck Foundation.  NIR data was
acquired by the authors as visiting astronomers at the Infrared
Telescope Facility, which is operated by the University of Hawaii
under Cooperative Agreement no. NCC 5-538 with the National
Aeronautics and Space Administration, Science Mission Directorate,
Planetary Astronomy Program.  The work of D.V. was supported by the
Czech Grant Agency (grant 205/08/0064) and the Research Program
MSM0021620860 of the Czech Ministry of Education.  We recognize the
cultural role and reverence that the summit of Mauna Kea has within
the indigenous Hawaiian community and appreciate the opportunity to
observe from this mountain.

\end{document}